\documentclass[10pt]{article}

\pdfoutput=1 
\usepackage{./aux/jheppub}
\usepackage[utf8]{inputenc}
\usepackage{graphicx,tikz}
\usepackage{amsfonts,amsmath,amssymb}
\usepackage{array,setspace,mathrsfs,yfonts,dsfont,bbm,colonequals,amscd,mathtools}
\usepackage{relsize,suffix,cancel,bbm,capt-of,upgreek,verbatim}

\newcommand{\be}{\begin{eqnarray}}
\newcommand{\ee}{\end{eqnarray}}
\newcommand{\bea}{\begin{eqnarray}}
\newcommand{\eea}{\end{eqnarray}}

\newcommand{\bn}{\begin{enumerate}}
\newcommand{\en}{\end{enumerate}}

\def\dim{{\rm dim}}

\def\fe{\mathfrak{e}}
\def\ff{\mathfrak{f}}
\def\fd{\mathfrak{d}}
\def\fg{\mathfrak{g}}
\def\fa{\mathfrak{a}}

\def \mf{\mathfrak}

\def \eg{{\it e.g.}}
\def \ie{{\it i.e.}}

\def\slice#1{\wr#1}

\usepackage{diagbox}

\numberwithin{equation}{section}

\def\NO#1#2{{\rm NO}(#1,#2)}

\def\bea{\begin{eqnarray}}
\def\eea{\end{eqnarray}}

\DeclarePairedDelimiterX\MeijerM[3]{\lparen}{\rparen}%
{\begin{smallmatrix}#1 \\ #2\end{smallmatrix}\delimsize\vert\,#3}

\newcommand\MeijerG[8][]{%
  G^{\,#2,#3}_{#4,#5}\MeijerM[#1]{#6}{#7}{#8}}

\WithSuffix\newcommand\MeijerG*[7]{%
  G^{\,#1,#2}_{#3,#4}\MeijerM*{#5}{#6}{#7}}

\def\fg{\mathfrak{g}}

\def \beg#1{\begin{#1}} 
\def \be{\beg{equation}}
\def \bea{\beg{eqnarray}}
\def \eea{\end{eqnarray}}
\def \ee{\end{equation}}

\def \gf{\mf{g}}

\def \slf{\mf{sl}}
\def \suf{\mf{su}}

\def \restr#1#2{{\left.\kern-\nulldelimiterspace#1\vphantom{\big|}\right|_{#2}}}

\def \mf{\mathfrak}

\def \eg{{\it e.g.}}
\def \ie{{\it i.e.}}

\def \Cb{\mathbb{C}}

\def \Zb{\mathbb{Z}}

\def \FF{\mathcal{F}}

\def \II{\mathcal{I}}

\def \MM{\mathcal{M}}
\def \NN{\mathcal{N}}

\def \QQ{\mathcal{Q}}
\def \RR{\mathcal{R}}
\def \SS{\mathcal{S}}
\def \TT{\mathcal{T}}

\def \VV{\mathcal{V}}
\def \WW{\mathcal{W}}

\title{Free Field Realizations from the Higgs Branch}

\preprint{YITP-SB-19-6}

\author[a,b]{Christopher Beem,\!}
\author[a]{Carlo Meneghelli,\!}
\author[c, d]{and Leonardo Rastelli}

\affiliation[a]{Mathematical Institute, University of Oxford, Woodstock Road, Oxford, OX2 6GG, United Kingdom}
\affiliation[b]{St. John's College, University of Oxford, St. Giles Road, Oxford, OX1 3JP, United Kingdom}
\affiliation[c]{Yang Institute for Theoretical Physics, State University of New York, Stony Brook, NY 11794-3840, USA}
\affiliation[d]{CERN, Theoretical Physics Department, 1211 Geneva 23, Switzerland}

\emailAdd{christopher.beem@maths.ox.ac.uk}
\emailAdd{carlo.meneghelli@maths.ox.ac.uk}
\emailAdd{leonardo.rastelli@stonybrook.edu}

\abstract{
We present free field realizations for the associated vertex operator algebras of a number of four-dimensional $\NN=2$ superconformal field theories. Our constructions utilize an exceptionally small set of chiral bosons whose number matches the complex dimensionality of the Higgs branch of the superconformal field theory. In the case of theories whose Higgs branches support additional degrees of freedom (free vector multiplets or decoupled interacting SCFTs), the corresponding ``free field realizations'' include additional ingredients: symplectic fermions in the case of vector multiplets and a $C_2$ co-finite VOA in the case of a residual interacting SCFT. The resulting picture is that the associated VOA can be constructed from the Higgs branch effective theory via free field realization. Our constructions also provide a natural realization of the $R$-filtration of the associated VOA.
}


\begin{document} 

\maketitle
\flushbottom



\vspace{-20pt}

\section{Introduction}
\label{sec:intro}

To any four-dimensional $\NN=2$ superconformal field theory (SCFT) one can canonically associate a two-dimensional vertex operator algebra (VOA) \cite{Beem:2013sza},
\begin{equation}
\label{eqn:correspondence}
\chi: \;\; {\rm 4d \;\; \NN=2 \; \; SCFT ~\longrightarrow~ VOA}~.
\end{equation}
The VOA $\chi[\TT]$ arises as a cohomological reduction of the full local OPE algebra of a four-dimensional theory $\TT$ with respect to a certain nilpotent supercharge.\footnote{The supercharge takes the schematic form $\QQ + \SS$, where $\QQ$ and $\SS$ denote a Poincar\'e and a conformal supercharge, respectively.}

This correspondence between four-dimensional SCFTs and two-dimensional VOAs illuminates both sides. The VOA computes a rich class of observables of the SCFT, which would be difficult or even impossible to obtain by other methods, notably when the SCFT has no known Lagrangian description. In the opposite direction, four-dimensional physics expectations lead to interesting conjectures for some large classes of VOAs. A prime example is the predicted existence of a two-dimensional topological quantum field theory ``valued in VOAs'', which was first inferred from the class $\SS$ duality web of four-dimensional SCFTs \cite{Beem:2014rza}, and has subsequently been given a rigorous mathematical construction \cite{Arakawa:2018egx}. Finally, one may use this $4d/2d$ correspondence as an organizing principle for the whole landscape of $\NN=2$ superconformal field theories. Indeed, a natural ambition is to develop a classification program for $\NN=2$ SCFTs with the associated VOAs playing a foundational role. A first set of results in this direction are certain universal bounds on four-dimensional central charges \cite{Beem:2013sza, Liendo:2015ofa, Lemos:2015orc, Beem:2018duj}, derived by combining the analytic power of the VOA structure with certain positivity constraints related to four-dimensional unitarity. 

\smallskip

The vertex algebra $\chi[\TT]$ represents an intricate shadow of the full SCFT $\TT$. It is independent of exactly marginal couplings, and as such it captures ``protected'' data associated to the whole conformal manifold in which a given SCFT lives. There are numerous indications that $\chi[\TT]$ is deeply connected with the physics of the Higgs branch of vacua of $\TT$, though the full extent of the connection remains somewhat elusive. A remarkable fact, observed in many examples and conjectured to be universally true \cite{Beem:2017ooy}, is that one can recover directly from $\chi[\TT]$ the Higgs branch $\MM_H[\TT]$, as a holomorphic symplectic variety,
\begin{equation}
\label{eqn:M=X}
\MM_H [ \TT]= X_{\chi [\TT]}~,
\end{equation}
where $X_\VV$ denotes the \emph{associated variety} of a VOA $\VV$ \cite{Arakawa:2010ni}. This conjecture has deep consequences for the structure of $\chi[\TT]$,%
\footnote{In particular, the fact that $X_{\chi[\TT]}$ must be symplectic implies that $\chi[\TT]$ exhibits remarkable modular properties -- its vacuum character must obey a monic linear modular differential equation \cite{Beem:2017ooy}, even though in general $\chi[\TT]$ is irrational.} and it is one of the strongest indications to date that the VOAs arising in connection with four-dimensional SCFTs are generally of a very special type.%
\footnote{For example, for \eqref{eqn:M=X} to hold, $\chi[\TT]$ must possess certain singular vectors, which among other things encode relations of the Higgs chiral ring.} Characterizing precisely what properties these VOAs must possess, and determining in exactly what way the associated VOA encodes the physics of the Higgs branch, are two major outstanding problems in this area.
 
In this paper we begin to address both of these problems. Though a comprehensive theory remains at large, we observe a uniform pattern in a large class of examples that allows us to begin extracting a general picture. We find that VOAs associated to four-dimensional SCFTs admit particularly nice \emph{free field realizations}. Free field realizations are a familiar and useful tool from two-dimensional CFT -- standard examples being the Wakimoto representation of affine current algebras at general level and the vertex operator representation of affine current algebras at level one. In contrast to these general and well-known constructions, an important feature of the free field realizations that arise in the present context is that they realize the \emph{simple quotient} of the relevant VOA, \ie, all null vectors vanish identically when expressed in terms of the free fields. This is a very delicate property, as the VOAs in question occur at specific values of the central charges and other parameters such that there is generally an array of non-trivial singular vectors.%
\footnote{VOAs associated to four-dimensional SCFTs are typically ``non-deformable'', \ie, they exist only for isolated values of the central charge. When they happen to belong to continuous families, they correspond to four-dimensional theories only at isolated points of their parameter space.} A related fact is that our free field constructions are highly economical, in that they use a smaller number of chiral bosons than would be required to realize the VOAs in question at generic values of their parameters. This is analogous to the well-known fact that critical-level affine Kac-Moody VOAs can be realized using fewer chiral bosons than their general-level cousins.

Strikingly, we find that the free field realization of $\chi[\TT]$ mirrors the effective field theory (EFT) description of $\TT$ on the Higgs branch of its moduli space of vacua. Examples can accordingly be organized by the complexity of their EFT/free field realization roughly as follows:
\begin{enumerate}
\item[(i)] In the simplest cases, the low energy degrees of freedom at a generic point of the Higgs branch consist of $n$ free, massless half-hypermultiplets, with $n = {\dim}_{\Cb} {\MM}_H[ \TT]$. In these examples, where the Higgs branch theory is purely geometric, the free field realization for $\chi[\TT]$ is given in terms of $n$ chiral bosons, in one-to-one correspondence with the $n$ half-hypermultiplets describing the Higgs branch EFT. Furthermore, the expressions for the generators of the VOA in terms of these free fields are determined by the structure of the Higgs branch as an algebraic variety.
\item[(ii)] At the next layer of complexity we encounter theories for which an abelian gauge group $U(1)^k$ remains unbroken at generic points of the Higgs branch, such that $k$ free vector multiplets survive in the low energy EFT. In such cases, we find that the free field realization must include (in addition to the chiral bosons encoding the Higgs branch geometry) $k$ symplectic fermion pairs.
\item[(iii)] Finally, the most general (though perhaps least familiar) situation is when, in addition to free fields, a decoupled interacting SCFT ${\TT}_{\rm IR}$ also survives at generic Higgs branch points. In such cases, we find that $\chi[\TT]$ admits a ``generalized free field'' construction of in terms of chiral bosons and the VOA $\chi[\TT_{\rm IR}]$, viewed as an irreducible building block. As we are considering the EFT of ${\TT}$ at a \emph{generic} point of its Higgs branch, the theory $\TT_{\rm IR}$ will itself have a trivial Higgs branch. Thus, according to the relation \eqref{eqn:M=X}, $\chi[\TT_{\rm IR}]$ must have a zero-dimensional associated variety, which in technical terms means that it will be $C_2$-cofinite. The case described above in item $(ii)$ can be thought of as a special case of this scenario, where the $C_2$-cofinite theory is a free theory of symplectic fermions.
\end{enumerate}
The existence and form of our free field realizations suggests that the VOAs arising from four-dimensional SCFTs are even more deeply connected to the underlying Higgs branch geometry than was previously understood, and in fact, may ultimately be recoverable from the Higgs branch (decorated with the data of the EFT $\TT_{\rm IR}$ living at the generic Higgs branch point).

An additional benefit of our free field constructions is that, by virtue of their transparent connection to Higgs branch geometry, they come equipped with a natural filtration that can be (conjecturally) identified with the filtration by $\suf(2)_R$ charge described in detail in \cite{Beem:2017ooy}. As we will recall in Section \ref{sec:preliminaries}, the $R$-filtration is an important structure that is inherited by the associated VOA of a four-dimensional SCFT that has hitherto not been given an intrinsic VOA definition. However, it is a key ingredient in the detailed identification of two- and four-dimensional operators, and so in the extraction of implications of four-dimensional unitarity for the structure of associated VOAs. By geometrizing the associated VOA in the manner studied in this work, we can plausibly endow the associated VOA with an intrinsic $R$-filtration.

In this article, we describe in detail a set of examples that represent arguably the simplest instances of each of the three categories outlined above. An additional simplifying feature that is present in all of the examples discussed here is that the only singularity of the Higgs branch is at the origin, where the original SCFT lives. The more general case involves a Higgs branch that is a stratified symplectic space, with various singular sub-strata supporting intermediate interacting low-energy degrees of freedom between those of $\TT$ and those of $\TT_{\rm IR}$. Additionally, in all cases described here the Higgs branch is spanned by Goldstone bosons, which is a special feature not enjoyed in generic cases. More elaborate examples that include nontrivial stratifications and flat directions not related to Goldstone degrees of freedom will appear in a future publication \cite{BMRToAppear}. 

The organization of this article is as follows. In Section \ref{sec:preliminaries} we recall the essential aspects of the correspondence \eqref{eqn:correspondence}, though an unfamiliar reader may need to consult the previous literature, especially \cite{Beem:2013sza} and \cite{Beem:2017ooy}. In Section \ref{sec:Deligne} we describe free field realizations of the type described above for the VOAs associated to the Deligne-Cvitanovi\'c exceptional series of simple Lie algebras. These are all examples where the Higgs branch theory is purely geometric, so the free field constructions use only chiral bosons. In Section \ref{sec_N4} we consider the case of $\NN=4$ super Yang-Mills theory with $SU(2)$ gauge group. In this case there is a single free abelian vector multiplet present at generic points of the Higgs branch, and accordingly the free field realization involves an extra symplectic fermion pair. In Section \ref{sec_AD} we consider the $(A_1,D_{2n+1})$ series of Argyres-Douglas SCFTs, for which the Higgs branch EFT includes a factor that is identified with the $(A_1,A_{2n-2})$ Argyres-Douglas SCFT. The ``free field realizations'' for these theories involve factors that are $C_2$ co-finite Virasoro VOAs, which are the associated VOAs for the $(A_1,A_{even})$ Argyres-Douglas theories. The constructions of this section have appeared before for different reasons in work of Adamovic \cite{Adamovic:2017}. In Section \ref{sec_filtration} we return to the issue of the $R$-filtration of \cite{Beem:2017ooy} and we conjecture a prescription for identifying the $R$-filtration using our free field constructions, and subject our conjecture to some basic checks. We conclude with some remarks about promising directions for future investigation in Section \ref{sec_conclusions}.


\section{Geometric VOA preliminaries}
\label{sec:preliminaries}

In this section we record some foundational aspects of the SCFT/VOA correspondence that will be relevant to our study below. This quick review is not meant to be self-contained -- we refer to the original paper \cite{Beem:2013sza} for the basic VOA construction and its more detailed features. We emphasize here the aspects of the correspondence that connect most directly with Higgs branch physics of the associated SCFT \cite{Beem:2014rza, Beem:2017ooy}.

\subsection{Schur operators and the \texorpdfstring{$R$}{R}-filtration}
\label{subsec:R}

We first recall some basic notations. We shall denote by $(E, j_1, j_2, R, r)$ charges of local operators under the five Cartan generators of the four-dimensional superconformal algebra $\mathfrak{su}(2, 2|2)$, with $E$ denoting the scaling dimension, $(j_1, j_2)$ the Lorentz spins, and $(R,r)$ the charges under Cartan generators of the $\mathfrak{su}(2)_R \oplus \mathfrak{u}(1)_r$ R-symmetry. The underlying vector space $\VV$ of the VOA $\chi[{\TT}]$ is isomorphic to the space of \emph{Schur operators} of the parent four-dimensional SCFT $\TT$. By definition, Schur operators are those local operators that obey
\begin{equation}
E = 2 R + j_1 + j_2~,\qquad r = j_2 - j_1~,
\end{equation}
where the second relation is a consequence of the first if one assumes (as we always shall) that the four-dimensional theory is unitary. Schur operators are thus characterized by the three quantum numbers $(E, R, r)$. Under the SCFT/VOA correspondence, each Schur operator gives rise to a state in $\VV$, with holomorphic scaling dimension 
\begin{equation}
h = E- R = R + j_1 +j_2~.
\end{equation}
The triple grading of the space of Schur operators descends to a triple grading of the VOA vector space,
\begin{equation}
{\cal V} = \bigoplus_{h, R, r}  {\cal V}_{h, R, r}~.
\end{equation}
It is important to recognize that while the $h$ and $r$ gradings are natural from the viewpoint of the VOA structure, the $R$ grading is not. For instance, the normally ordered product preserves $h$ and $r$,\footnote{The preservation of $h$ is in the usual sense of conformal VOAs, where the spacetime coordinate $z$ has scaling dimension minus one in the OPE.} but violates $R$. However, the specifics of the cohomological construction of \cite{Beem:2013sza} imply that $R$-charge can at most {\it decrease} under normally-ordered multiplication,
\begin{equation}
\NO{\VV_{h_1,R_1,r_1}}{\VV_{h_2,R_2,r_2}}\subseteq \bigoplus_{k\geqslant0} \VV_{h_1+h_2,R_1+R_2-k,r_1+r_2}~.
\end{equation}
Consequently, there is a \emph{filtration} by $R$ that is preserved by the normally-ordered product. That is, if we define,
\begin{equation}\label{eq:R-filtration}
\FF_{h,R,r}=\bigoplus_{k\geqslant 0}\VV_{h,R-k,r}~,
\end{equation}
then we have
\begin{equation}\label{eq:R-filtered-product}
{\rm NO}(\FF_{h_1,R_1,r_1},\FF_{h_2,R_2,r_2})\subseteq \FF_{h_1+h_2,R_1+R_2,r_1+r_2}~.
\end{equation}
In addition, a secondary bracket operation defined by taking the simple pole in the OPE of two operators obeys
\begin{equation}\label{eq:R-filtered-bracket}
\{\FF_{h,R,r},\FF_{h',R',r'}\}\subseteq \FF_{h+h'-1,R+R'-1,r+r'}~,
\end{equation}
so the bracket is filtered of tri-degree $(-1,-1,0)$.

VOAs that arise from four-dimensional SCFTs are necessarily equipped with an $R$-filtration as we have just described. Furthermore, knowledge of the $R$-filtration is a prerequisite for a detailed understanding of the identification between two- and four-dimensional operators, which is turn required in many applications of VOA technology to four-dimensional physics. In particular, the $R$-filtration is needed if one aims to impose the full constraints of four-dimensional unitarity beyond simple systems of correlators such as those studied in \cite{Beem:2013sza, Liendo:2015ofa, Lemos:2015orc, Beem:2018duj}. 

To date, it has not been clearly understood whether the $R$-filtration can be given a description that is \emph{intrinsic} to the VOA (though see \cite{Song:2016yfd} for some progress in this direction). Namely, given an abstract presentation of the VOA, there is no known general recipe to define the $R$-filtration. A remarkable and important feature of the free field realizations described in this paper (as well as those in \cite{Bonetti:2018fqz}) is that they come equipped with a natural (and intrinsic) filtration with the right properties to be identified with the $R$-filtration inherited from four-dimensions. Indeed, by anticipating this application, the form of our free field constructions can be partly motivated by the requirement that they give rise to the right commutative vertex algebra structure at the associated-graded level.

\subsection{Higgs branch and Hall-Littlewood operators as (strong) generators}
\label{subsec:strong_generators}

A VOA is called \emph{strongly finitely generated} if it possesses a finite collection of \emph{strong generators} such that normally ordered products of those strong generators and their derivatives span the space of states of the VOA. If a VOA is strongly finitely generated, then it (or more precisely, its simple quotient) is completely characterized by the singular terms in the OPEs between strong generators.

It is generally thought that the VOAs arising from four-dimensional SCFTs (with finite central charge) are all strongly finitely generated, though the strong generators have not been characterized entirely in four-dimensional terms. However, it is known that the \emph{generators} of the Higgs branch chiral ring (\ie, those operators obeying the half-BPS condition $E = 2R$) necessarily give rise to strong generators of the VOA. Furthermore, when the Hall-Littlewood chiral ring is different from the Higgs branch chiral ring, then additional generators of the Hall-Littlewood chiral ring (which obey the more general BPS condition $E = 2R+j_1$) also correspond to strong VOA generators. There may also be additional strong generators: for example, if the central charge in an interacting SCFT does not take the Sugawara value for the flavor symmetry of the theory, then the stress tensor will always be an additional strong generator of the associated VOA.

A weaker condition than being strongly finitely generated is for a VOA to be finitely generated. This is the case when there exists a finite number of \emph{generators} of the VOA such that the full space of states can be constructed from the action of the modes of those generators on the vacuum -- this include the action of positive modes, \ie, one has access to operators that appear in the singular OPEs of the generators, not just their normally ordered products.

An attractive possibility that, to the authors' knowledge, is compatible with all currently known examples, is that in theories with nontrivial Higgs branches the VOA operators arising from generators of the Hall-Littlewood chiral (and anti-chiral) ring \emph{generate} the associated VOA, even when they do not strongly generate it. As we will see in subsequent sections, in the context of free field realizations a knowledge of the generators of the VOA is quite powerful. If one has a construction of a set of generators as a elements of a free field VOA, the full VOA is then determined automatically. This perspective will be further explored in \cite{BMRToAppear}.

\subsection{Associated varieties, quasi-lisse vertex algebras, and the Higgs branch}
\label{sec:assoc_var}

Given any VOA $\VV$, one may define the vector subspace
\begin{equation}
C_2(\VV) = {\rm span}\left\{a_{-h_a-1}b~,~a, b\in\VV\right\}~.
\end{equation}
Rearrangement formulae for normally ordered products imply that this space is spanned by all possible normally ordered products that involve a derivative of a strong generator in any position. The \emph{commutative Zhu algebra} can then be defined as the vector space quotient,
\begin{equation}
\RR_\VV \cong \VV/C_2(\VV)~.
\end{equation}
Multiplication in $\RR_\VV$ is induced by the normally-ordered product, but in the quotient space, this product becomes (super) commutative and associative. Additionally, at the level of the quotient space, the secondary bracket mentioned above obeys the rules of a Poisson bracket with respect to multiplication. Thus $\RR_\VV$ is a Poisson algebra, graded by conformal dimension.

Arakawa has defined the associated variety of a VOA according to
\begin{equation}
X_\VV = {\rm mSpec}\left(\RR_\VV\right)~.
\end{equation}
Roughly, this means that $X_\VV$ is an affine algebraic variety defined by polynomial equations that are the images singular vectors of $\VV$ in the quotient $\RR_\VV$. In practice, the ring $\RR_\VV$ may include nilpotent elements, and the associated variety is defined by the reduction of $\RR_\VV$ in which one further quotients by the nilradical.

It was conjectured in \cite{Beem:2017ooy}, and has since been confirmed in a large number of examples, that for VOAs associated to four-dimensional SCFTs the associated variety is equivalent to the Higgs branch. In \cite{Beem:2017ooy} it was explained that this conjecture requires the vanishing of a certain ideal in $\RR_\VV$ whose definition refers to the $R$-filtration on $\VV$. Subject to this conjecture, one learns that the associated VOAs of four-dimensional SCFTs have associated varieties that are \emph{symplectic}.\footnote{More generally, the associated variety may be a stratified symplectic space with a finite number of symplectic leaves.} Such VOAs have been dubbed \emph{quasi-lisse} by Arakawa and Kawasetsu \cite{Arakawa:2016hkg}. This property suggests that in some sense, these VOAs should be more geometric than a generic VOA, with the finite-dimensional associated variety playing a central role.

\section{\texorpdfstring{Deligne-Cvitanovi\'c}{Deligne-Cvitanovi\'c} exceptional series}
\label{sec:Deligne}

We now turn to our first set of examples: the (simple) affine Kac-Moody vertex algebras associated to the Deligne-Cvitanovi\'c (DC) series of finite-dimensional, simple Lie algebras\footnote{The case of $\fa_0$ is special and is identified with the Virasoro VOA at central charge $c=-\frac{22}{5}$.}
\begin{equation}
\fa_0\,\subset\,
\fa_1\,\subset\,
\fa_2\,\subset\,
\fg_2\,\subset\,
\fd_4\,\subset\,
\ff_4\,\subset\,
\fe_6\,\subset\,
\fe_7\,\subset\,
\fe_8~,
\end{equation}
at level 
\begin{equation}
k\,=\,-\frac{h^\vee}{6}-1~,
\end{equation}
where $h^\vee$ denotes the dual Coxeter number. This family of VOAs has been singled out in \cite{Beem:2013sza} as the set of theories that simultaneously saturate four-dimensional unitarity bounds\footnote{For all cases except for $\fg_2$ and $\ff_4$, these VOAs are related to the theory of a single $D3$ brane probing a singular fiber in an $F$-theory compactification on a $K3$ surface. The four dimensional interpretation, if any, of the $\fg_2$ and $\ff_4$ entries remains elusive.} for the central charge $c_{4d}$ and the flavor central charge $k_{4d}$. Additionally, it has been proved in \cite{Arakawa:2015jya} that for this family of VOAs the associated variety is the closure of the minimal nilpotent orbit $\mathbb{O}_{\text{min}}(\mathfrak{g})$ of the corresponding Lie algebra $\mathfrak{g}$.\footnote{The VOAs corresponding to the $\fd_4$, $\fe_6$ and $\fe_7$ entries have also been recently considered in \cite{Dedushenko:2017osi,Eager:2019zrc}. These works suggest that these VOAs can be described using sigma models/curved $\beta\gamma$ systems with singular targets. It would be interesting to clarify the relation of these works to our present construction.}

We will present free field realizations of these affine Kac-Moody VOAs that are specialized to the levels in question. As we have anticipated in the introduction, our free field realizations will utilize $n=\dim_{\mathbb{C}}\mathcal{M}_{H}=2(h^{\vee}-1)$ chiral bosons, which we take as being in one-to-one correspondence with the half-hypermultiplets appearing in the Higgs branch effective field theory. In all cases there will be one pair of chiral bosons that is treated differently from the remaining $n-2$, which are combined into $h^{\vee}-2$ copies of the $(\beta,\gamma)$ VOA with conformal weights $(\tfrac{1}{2},\tfrac{1}{2})$ (\ie, $h^\vee-2$ symplectic bosons).

This asymmetry can be briefly understood as follows (see also \cite{Shimizu:2017kzs}). The space $\mathcal{M}_H=\overline{\mathbb{O}_{\text{min}}(\mathfrak{g})}$, as a hyperk\"ahler manifold, enjoys an $SU(2)_R\times G$ isometry group, where $\mathfrak{g}=\text{Lie}(G)$. At any point on $\mathcal{M}_H$ different from the origin this symmetry is broken spontaneously to $SU(2)_{\bar{R}}\times G^{\natural}$ where $SU(2)_{\bar{R}}=\text{diag}\left(SU(2)_R\times SU(2)_{\theta}\right)$ and the subgroup $SU(2)_{\theta}\times G^{\natural}\subset G$ is described in more detail in Section \ref{sec:minNILpatch}. The tangent space at the given point organizes into a representation of the residual symmetry as $(\mathbf{1} \oplus \mathbf{3}, \mathbf{1})\oplus (\mathbf{2} ,\mathfrak{R})$, where $\dim\,\mathfrak{R}=2(h^{\vee}-2)$.\footnote{In the description of $\mathcal{M}_H$ as a holomorphic symplectic variety, only the abelian subgroup $SO(2)_{\bar{R}}\subset SU(2)_{\bar{R}}$ generated by $\bar{R}=R-\tfrac{1}{2}h_{\theta}$ (with $h_{\theta}$ the Cartan generator of $\mathfrak{sl}(2)_{\theta}$) is visible. The (holomorphic) tangent space decomposes as $\mathbf{1}_0\oplus \mathbf{1}_1\oplus \mathfrak{R}_{\frac{1}{2}}$, where the suffix indicates the $\bar{R}$-charge.} The four real directions that are neutral with respect to $G^{\natural}$ are associated to the two distinguished chiral bosons mentioned above, while the remaining directions that transform in the representation $\mathfrak{R}$ give rise to symplectic bosons.

\subsection{The \texorpdfstring{$\fa_1$}{a(1)} free field realization}
\label{subsec:A1_Deligne}

To demonstrate the general structure of our construction, we first consider in detail the simplest (geometric) entry of the Deligne-Cvitanovi\'c exceptional series: the affine Kac-Moody VOA $V_{-\frac43}(\slf(2))$, which is given in a standard basis $\{e(z),f(z),h(z)\}$ by the OPEs
\begin{subequations}\label{affineSL2}
\begin{alignat}{2}
& e(z)e(w)\sim 0~,                                  \qquad && f(z)f(w)\sim 0~,\\
& h(z)e(w)\sim \frac{2\,e(w)}{z-w}~,                \qquad && h(z)f(w)\sim \frac{-2\,f(w)}{z-w}~,\\
& e(z)f(w)\sim \frac{k}{(z-w)^2}+\frac{h(w)}{z-w}~, \qquad && h(z)h(w)\sim\frac{2k}{(z-w)^2}~,
\end{alignat}
\end{subequations}
with $k=-4/3$. To motivate our free field construction, we will begin with a detailed discussion of the associated variety of this VOA as a holomorphic-symplectic variety. Our approach will be to ``approximate'' the associated variety with the cotangent bundle $T^\ast\mathbb{C}^\ast$, meaning we will find a (Zariski) open, dense subset that is equivalent to this cotangent bundle. We will think of the coordinates of this cotangent bundle as finite-dimensional, classical analogues of the free fields that we will use in our construction, and we can realize the Higgs chiral ring in terms of them. We then look for an affine uplift, which singles out three possible (and interesting) values for the level $k$.

\medskip

The Higgs branch/associated variety in this case is the closure of the unique nilpotent orbit of $\slf(2)$, \ie, the full nilcone,
\begin{equation}
\mathcal{M}_{\text{Higgs}} = \overline{\mathbb{O}_{\text{nil}}(\slf(2))} = \big{\{} x\in \slf(2)\big{|}\,x^2=0\big{\}}~.
\end{equation}
This has the structure of an algebraic variety when written in terms of the $\slf(2)$-valued function\footnote{We will freely identify $\gf$ with $\gf^\ast$ using the invariant Killing form.}
\begin{equation}\label{xxiszeroSL2}
x\,=\,
\begin{pmatrix}
\mathsf{p} & \mathsf{e} \\
\mathsf{f} & -\mathsf{p}
\end{pmatrix}~,
\qquad
x^2=\begin{pmatrix}
\mathsf{p}^2+\mathsf{e}\,\mathsf{f} &0 \\
0 &\mathsf{p}^2+\mathsf{e}\,\mathsf{f}
\end{pmatrix}\,
\stackrel{!}{=}\,0
~.
\end{equation}
From the second equation one recognizes the alternate description $\mathcal{M}_{\text{Higgs}}=\mathbb{C}^2/\mathbb{Z}_2$. The holomorphic symplectic structure is encoded in the Poisson brackets of the component functions of $x$, and these are just given by the $\slf(2)$ Lie algebra,
\begin{equation}\label{PoissonSL2}
\{\mathsf{h}, \mathsf{e}\}=+2\,\mathsf{e}~,
\qquad
\{\mathsf{h}, \mathsf{f}\}=-2\,\mathsf{f}~,
\qquad
\{\mathsf{e}, \mathsf{f}\}=\mathsf{h}~,
\end{equation}
where we have defined $\mathsf{h}=2\mathsf{p}$. In terms of this basis for the generators of the coordinate ring of $\MM_{\rm Higgs}$, we can perform a nilpotent Higgsing by setting $\langle\mathsf{e}\rangle=A\neq0$. Because the Higgs branch is a just a nilpotent orbit, the most general point away from the origin of $\MM_{\rm Higgs}$ takes this form after a change of basis by conjugation. We can now consider the open patch $\{\mathsf{e}\neq0\}\subset\MM_{\rm Higgs}$. In this patch, the relation \eqref{xxiszeroSL2} can simply be solved by setting
\begin{equation}\label{fpatch}
\mathsf{f}=-\mathsf{e}^{-1}\,\mathsf{p}^2~.
\end{equation}
Thus in this patch, there are coordinates $\mathsf{e}\in\Cb^\ast$ and $\mathsf{p}\in\Cb$, with Poisson bracket
\begin{equation}\label{finsl2}
\{\mathsf{p},\mathsf{e}\}=\mathsf{e}~,\qquad \{\mathsf{p},\mathsf{e^{-1}}\}=-\mathsf{e^{-1}}~.
\end{equation}
This just describes the cotangent bundle $T^\ast\mathbb{C}^\ast$ with its canonical symplectic form. Moreover, this patch is invariant under the scaling $\Cb^\ast$ action on $\MM_{\rm Higgs}$, which gives the simple $R$-charge assignments $R[\mathsf{p}]=R[\mathsf{e}]=1$, $R[\mathsf{e^{-1}}]=-1$.

We will look for a free field realization that ``affinizes'' this simple, finite-dimensional geometric construction. To do so, we introduce two chiral bosons $\delta(z),\varphi(z)$ with OPEs
\begin{equation}
\label{deltaphiOPE}
\delta(z_1)\delta(z_2)\sim \langle \delta,\delta\rangle\,\log z_{12}~,
\qquad
\varphi(z_1)\varphi(z_2)\sim \langle \varphi,\varphi\rangle\,\log z_{12}~,
\qquad
\delta(z_1)\varphi(z_2)\sim0~,
\end{equation}
where $z_{12}=z_1-z_2$. The VOA avatar of the $\Cb^\ast$-valued $\mathsf{e}$ in our finite-dimensional model is the exponentiated chiral boson
\begin{equation}\label{eSL2}
e(z)\,=\,e^{\delta(z)+\varphi(z)}~,
\end{equation}
which will eventually be identified with the corresponding generator of the affine Kac-Moody algebra VOA. In order for this identification to be reasonable, we have to verify that the OPE of $e(z)$ with itself is regular, which in turn requires that
\begin{equation}
\langle \varphi,\varphi\rangle=-\langle \delta,\delta\rangle~,
\end{equation}
so the combination $\delta(z)+\varphi(z)$ appearing in the exponential is a null direction. 

A natural first proposal for an affine uplift of the conjugate variable $\mathsf{p}$ is then given by the complementary null chiral boson,
\begin{equation}\label{pzdef}
 p(z)=\tfrac{\,1}{2\,\langle \delta,\delta \rangle}\partial(\delta(z)- \varphi(z))~.
\end{equation} 
This choice (and normalization) is motivated by the requirement that the simple poles in the OPEs among $e(z)$ and $p(z)$ provide reproduce the Poisson brackets \eqref{finsl2},
\begin{equation}\label{peOPE}
{\rm Res}_{z=w}\left(p(z)e(w)\right)\,=\,e(w)~,\qquad
{\rm Res}_{z=w}\left(p(z)p(w)\right)\,=\,{\rm Res}_{z=w}\left(e(z)e(w)\right)\,=\,0~.
\end{equation}
The operators $e(z)$ and $p(z)$ generate a subVOA of the full rank $(1,1)$ lattice VOA associated to our two chiral bosons, that we can write schematically as\footnote{In order to generate this VOA one needs to also invert $e(z)$. One then has, for example, $\partial(\delta+\varphi)=\NO{e^{-1}}{\partial e}$, where $\NO{-}{-}$ denotes conformal normal ordering.}
\begin{equation}\label{PilattexVOA}
\Pi\,\colonequals \,\bigoplus_{n=-\infty}^{\infty}\,\left(V_{\partial\varphi}\otimes V_{\partial\delta}\right)\,e^{n(\delta+\varphi)}
\simeq \bigoplus_{n=-\infty}^{\infty}\,\left(V_{p}\otimes V_{\partial(\delta+\varphi)} \right)\,e^{n(\delta+\varphi)}~,
\end{equation}
where $V_j,~j\in\{\partial\varphi,\partial\delta,p,\partial(\delta+\varphi)\}$ is the abelian affine current VOA associated with the current $j$.

The subVOA $\Pi$ admits a natural ascending filtration that, in anticipation of our identification with the $R$-filtration described in Section \ref{sec:preliminaries}, we will denote by $R$. We define\footnote{In equation \eqref{eq:filtration_def} we use the convention of nested conformal normal ordering $(A_1A_2\ldots A_{n-1}A_n) \colonequals (A_1(A_2(\cdots(A_{n-1}A_n)\cdots)))$. More generally, unless otherwise specified, products of free fields in this paper are assumed to be \emph{creation/annihilation normally ordered}.}
\begin{equation}\label{eq:filtration_def}
F_R\Pi \colonequals {\rm span}\left\{(\partial^{n_1}(\delta-\varphi)\cdots \partial^{n_k}(\delta-\varphi)\partial^{m_1}(\delta+\varphi)\cdots \partial^{m_\ell}(\delta-\varphi)e^{n(\delta+\varphi)})~,~k+n\leqslant R\right\}~,
\end{equation}
in terms of which we clearly have
\begin{equation}
\Pi = \bigcup_{i\in\mathbb{Z}}F_i\Pi~,\qquad F_i\Pi\subseteq F_j\Pi~,\quad i\leqslant j~.
\end{equation}
The components $F_{R}\Pi$ are infinite dimensional for any $R\,\in\mathbb{Z}$ and $R$ is unbounded below, but this will not cause us any real trouble since the VOA $\Pi$ admits two further gradings by conformal weight and by $U(1)$ charge relative to the affine current $\partial(\delta-\varphi)(z)$, which we denote by $h$ and $m$, respectively. It is then easy to see that the filtered subspaces $(F_R\Pi)_{h,m}$ with fixed values of $h$ and $m$ are finite-dimensional, and furthermore for fixed $h$ and $m$ the filtration truncates to the left (\ie, one has $(F_{R}\Pi)_{h,m}=\{0\}$ for $R<m$). 

This filtration enjoys good properties analogous to those attributed to the $R$-filtration in Section \ref{sec:preliminaries}: the normally ordered product and secondary bracket are compatible with the filtration as in \eqref{eq:R-filtered-product} and \eqref{eq:R-filtered-bracket}. As such, the associated graded with respect to this filtration is a commutative vertex algebra, which is generated by $e^{\delta+\varphi}$, $e^{-\delta-\varphi}$, and $\partial(\delta-\varphi)$ subject to the obvious relations under multiplication and differentiation.\footnote{Here we are overloading the notation for these operators, using the same notation in the associated graded ${\rm Gr}_R\Pi$ as in the VOA $\Pi$.} This commutative algebra has a natural geometric interpretation in terms of our ``approximation'' of the Higgs branch: it is the arc space $J_\infty(T^\ast\Cb^\ast)$.\footnote{See, for example, \cite{ArakawaMoreau:Arc} for a discussion of arc spaces in connection with quasi-Lisse vertex algebras.} The $R$-grading on this commutative vertex algebra is the one inherited from the $\Cb^\ast$ scaling symmetry of the underlying cotangent bundle, and we summarize the various gradings below in Table \ref{Tab:free_field_grading}.

\begingroup
\renewcommand{\arraystretch}{1.3}
\begin{center}
\begin{tabular}{| c|| c |c | c|c |c |c  | }
\hline
& $\partial(\delta-\varphi)$ & $e^{n(\delta+\varphi)}$
& $\partial(\delta+\varphi)$ &  $\partial$ \\\hline \hline
$R$  &$1$ & $n$ & $0$ & $0$  
\\\hline
$h$  &  $1$ &  $n$& $1$ & $1$ \\\hline
$m$  &  $0$ &  $n$& $0$ & $0$ \\\hline
\end{tabular}
\end{center}
\label{Tab:free_field_grading}
\captionof{table}{Grading of basic elements of the commutative vertex algebra ${\rm Gr}_R\Pi$. The $R$-filtration on $\Pi$ is easily read off based on the classical limit, with the important additional simplification that there are no nontrivial relations amongst operators in $\Pi$.}
\endgroup

\medskip

With these constructs in place, we can realize the $V_{-\frac43}(\slf(2))$ OPE in terms of our free fields. The generators $\{e(z),f(z),h(z)\}$ all correspond to Higgs branch chiral ring generators with $R=h=1$, and compatibility with the $R$-filtration will be a significant constraint on the form of our realization. The generator $e(z)$ was given in \eqref{eSL2} as $e(z)=\exp(\delta+\varphi)$. We note that $(F_1\Pi)_{1,1}$ is one dimensional and spanned by this exponential, so this is the only acceptable option.

Next we consider the Cartan generator $h(z)$. It has $U(1)$ charge $m=0$, so must live in $(F_1\Pi)_{1,0}$, which is two dimensional and spanned by the previously introduced $p(z)$ and the additional null boson $\partial(\delta+\varphi)$. To produce the correct $h\times e$ OPE we must then have $h(z)=2\,p(z)+\alpha\,\partial(\delta+\varphi)(z)$ for some numerical parameter $\alpha$. Notice that the ambiguity is in terms of an element of $F_0\Pi$. To determine $\alpha$ we compare to the $h\times h$ OPE in \eqref{affineSL2} and find the condition $\alpha=\tfrac{1}{2}k$. This implies that, up to a redefinition of $\delta$ and $\varphi$,\!\footnote{The redefinition of $\delta,\varphi$ is as follows:
\begin{alignat}{2}
\delta &\mapsto \delta^\prime &=&
\left(\tfrac{1}{2}-\tfrac{1}{k\langle \delta,\delta\rangle }\right)\delta+
\left(\tfrac{1}{2}+\tfrac{1}{k\langle \delta,\delta\rangle }\right)\varphi~,\\
\varphi  &\mapsto \varphi^\prime &=&
\left(\tfrac{1}{2}+\tfrac{1}{k\langle \delta,\delta\rangle }\right)\delta+
\left(\tfrac{1}{2}-\tfrac{1}{k\langle \delta,\delta\rangle }\right)\varphi~.
\end{alignat}
It is easy to check that $\langle \delta^\prime,\varphi^\prime\rangle=0$, $\langle \varphi^\prime,\varphi^\prime\rangle=-\langle \delta^\prime,\delta^\prime\rangle=\tfrac{2}{k}$, and $\delta+\varphi=\delta^\prime+\varphi^\prime$.} we have
\begin{equation}\label{htheta}
h(z)=k\,\partial\varphi(z)~,\qquad \langle\varphi,\varphi\rangle=-\langle \delta,\delta\rangle=\frac{2}{k}~.
\end{equation}
It remains to determine the most complicated generator: $f(z)$. $(F_1\Pi)_{1,-1}$ is five-dimensional, and by requiring the correct $h\times f$ and $e\times f$ OPEs we find that 
\begin{equation}\label{fSL2}
f(z)=-\left((\tfrac{k}{2}\partial \delta)^2-\tfrac{k(k+1)}{2}\, \partial^2\delta\right)e^{-(\delta+\varphi)}~,
\end{equation}
where we have left the $z$ dependence on the right hand side implicit. We note that up to terms that are sub-leading in the $R$-filtration, this matches the classical expression for the relevant element of the Higgs branch chiral ring expressed in cotangent bundle coordinates \eqref{fpatch}. 

At this point, all the generators of the affine VOA are determined, but we still need to verify regularity of the $f\times f$ OPE. This implies a constraint on the level $k$ of the VOA, that to this point we have treated as a general parameter,\footnote{It may be worth noting that for $k=-\tfrac{4}{3}$ one can write $f(z)=-\tfrac{2}{9}Q\cdot e^{\delta-\varphi}$ in terms of the screening charge $Q=\int e^{-2\delta}$.}
\begin{equation}\label{valuesofkA1}
k~\in~\{-2,-\tfrac{1}{2},-\tfrac{4}{3}\}~.
\end{equation}
The first two levels correspond to the critical level and the $\mathbb{Z}_2$ orbifold of the $\beta\gamma$ system respectively. For these levels the free-field realization presented above can be rewritten in terms of a $(\beta,\gamma)$ pair of conformal weights $(1,0)$ and $(\tfrac{1}{2},\tfrac{1}{2})$ respectively by undoing the relevant bosonization, with the critical-level case realizing the standard Wakimoto representation of the critical $\slf(2)$ VOA.

Some remarks are in order at this point. The free field realization given in \eqref{eSL2}, \eqref{htheta}, \eqref{fSL2} with $k=-\tfrac{4}{3}$ has been introduced by Adamovic in \cite{Adamovic:2004}.\footnote{This free field realization has in fact appeared previously in the work of \cite{Semikhatov:1993pr}.} Secondly, as was shown in \cite{Adamovic:2004}, this construction yields the simple quotient of the affine vertex algebra at $k=-4/3$, \emph{i.e.}, singular vector are identically zero. We note also that  this free field construction uses only two chiral bosons, as opposed to the three employed in the more familiar (and valid for general level) Wakimoto realization.

\subsection{The general construction}
\label{subsec:general_Deligne}

The construction of the previous section can be generalized to the full DC exceptional series. We will adopt a similar strategy to what was done above for the $\fa_1$ case: we consider a patch of the associated variety/Higgs branch that is identified by a choice of nilpotent Higgsing. In this patch, the component of the moment map that was turned on will be invertible (to be precise, we will consider the localization of the Higgs chiral ring with respect to the relevant component of the moment map). Algebraically the resulting patch will take the form $\Cb^\ast\times\Cb^{\dim_{\mathbb{C}}\MM_H-1}$, \ie, we will be able to explicitly solve the Higgs chiral ring relations in terms of $\dim_{\mathbb{C}}\MM_H$ generators (and the inverted moment map) that obey no additional relations amongst themselves.

In contrast to the $\fa_1$ case, the Poisson structure on this patch will not take an entirely obvious form. However, a modest redefinition will allow us to describe the same patch as (a quotient of) a cotangent bundle, with its canonical Poisson brackets. We use this cotangent approximation of the Higgs branch as the starting point for an affine uplift in terms of a lattice VOA and an appropriate number of $(\beta,\gamma)$ systems. The resulting expressions for the generators of the VOA are given in \eqref{eandeDeligne}, \eqref{htheta}, \eqref{gnatural}, \eqref{fthetaDELIGNE}, and \eqref{fADEL}.

Remarkably, we will see that the existence of an affine uplift of our set-up singles out the DC series at the correct levels, together with VOAs associated to discrete quotients of free theories, namely $V_{-\frac{1}{2}}(\mathfrak{sp}(2n))\simeq (\mathbb{V}_{\beta \gamma})^{n}/\mathbb{Z}_2$. The general construction presented in this section is new, but the case of $\gf=\fa_2$ has appeared previously (with slightly different notation) in \cite{Adamovic:2014lra}.

\begin{table}[t]
\centering
\renewcommand{\arraystretch}{1.4}
\begin{tabular}[t!]{|c | c | c | c  | c| c|c|} 
\hline
$\gf$ &  $\gf^{\natural}$ & $\mathfrak{R}$ & $h^{\vee}$& $k$ & $I_{\gf^{\natural} \hookrightarrow \gf}$ & $I_{\gf^{\natural} \hookrightarrow \mathfrak{sp}(\mathfrak{R})}$\\ 
\hline
\hline
$\fa_1$  & $-$ & $-$ & $2$ & $-\frac{4}{3}$& $-$&    $-$\\ 
\hline
$\fa_2$  & $\mathbb{C} h_{\perp}$ & $\mathbf{1}_+ \oplus \mathbf{1}_-$ & $3$ & $-\frac{3}{2}$  & $-$&$-$ \\ 
\hline
$\fg_2$  & $\fa_1$ & $\mathbf{4}$ & $4$ & $-\frac{5}{3}$& $3 $& $10$  \\ 
\hline
$\fd_4$  & $\fa_1\oplus \fa_1\oplus \fa_1$ & $(\mathbf{2},\mathbf{2},\mathbf{2})$ & $6$  & $-2$ & $1,1,1$ & $4,4,4$ \\ 
\hline
$\ff_4$  & $\mathfrak{c}_3$ & $\mathbf{14}'$ & $9$  & $-\frac{5}{2}$&$1$&   $5$\\ 
\hline
$\fe_6$  & $\fa_5$ & $\mathbf{20}$ & $12$  & $-3$& $1$ & $6$  \\ 
\hline
$\fe_7$  & $\fd_6$ & $\mathbf{32}$ & $18$  & $-4$& $1$&  $8$  \\ 
\hline
$\fe_8$  & $\fe_7$ & $\mathbf{56}$ & $30$  & $-6$ & $1$ &  $12$\\ 
\hline
\hline
$\fa_r$  & $\fa_{r-2}\oplus \mathbb{C} h_{\perp}$ & 
$(\mathbf{r-1})_+ \oplus (\overline{\mathbf{r-1}})_-$ & $r+1$  & $-1$ &  $1$ & $2$   \\ 
\hline
$\mathfrak{b}_r$  & $\fa_{1}\oplus 
\mathfrak{b}_{r-2}$ & $(\mathbf{2},\mathbf{2r-3})$ &  $2r-1$ & $-\tfrac{2r-3}{2},-2$ & $1,1$ &$2r-3,4$  \\ 
\hline
$\mathfrak{c}_r$  & $\mathfrak{c}_{r-1}$ & $\mathbf{2(r-1)}$ &   $r+1$ & $-\tfrac{1}{2}$ &  $1$ & $1$ \\ 
\hline
$\fd_r$  & $\fa_1\oplus \fd_{r-2}$ & $(\mathbf{2},\mathbf{2r-4})$ &   $2r-2$ &$-\tfrac{2r-4}{2},-2$ & $1,1$ & $2r-4,4$\\ 
\hline
\end{tabular}
\label{Table_Deligne}
\captionof{table}{Lie-algebraic data for finite-dimensional simple Lie algebras relevant to the structure of the minimal nilpotent orbit. The top portion of the table corresponds to the Deligne-Cvitanovi\'c exceptional series of Lie algebras. The entry for the level is obtained by applying \eqref{LevelkfromtheSPembedding} (to each simple factor) when $\gf\neq \fa_1, \fa_2$. See the appendix of \cite{Argyres:2007cn} for a useful review of embedding indices.}
\end{table}

\subsubsection*{Minimal nilpotent orbits and a cotangent patch}
\label{sec:minNILpatch}

Let us recall the algebraic and geometric structure of minimal nilpotent orbits, which are the associated varieties/Higgs branches for the theories in question. For a fixed choice of $\gf$, let $\theta$ be a choice of highest root and $\slf(2)_{\theta}=\langle e_{\theta},f_{\theta},h_{\theta} \rangle \subseteq \gf$ the associated $\slf(2)$ triple. The minimal nilpotent orbit of $\gf$ is then defined to be
\begin{equation}
\mathbb{O}_{\min}(\gf)= G.e_{\theta}~,\qquad \gf=\text{Lie}(G)~.
\end{equation}
Here we denote the commutant subalgebra of $\slf(2)_\theta$ in $\gf$ by $\gf^\natural = Com(\slf(2)_\theta,\gf)$. With respect to $\slf(2)_\theta\times\gf^\natural$, the Lie algebra $\gf$ decomposes according to
\begin{equation}\label{d_decomp_Deligne}
\gf= \left(\gf^{\natural}\oplus \slf_2\right)\oplus (\mathfrak{R},2)~.
\end{equation}
where $\mathfrak{R}$ is a specific quaternionic representation of $\gf^{\natural}$. We supply the data of $\gf^\natural$ and $\mathfrak{R}$ for all finite-dimensional, simple Lie algebras in Table~\ref{Table_Deligne}. It will also prove useful to decompose $\gf$ according to its $h_\theta$ weight spaces, whereupon we have a ``five-graded'' structure,
\begin{equation}\label{eq:five_grade_decomposition}
\gf = \left(\mathbb{C}f_{\theta}\right)_{-2}\oplus\left(\mathfrak{R}^-\right)_{-1}\oplus\left(\gf^{\natural}\oplus\mathbb{C}h_{\theta}\right)_0\oplus\left(\mathfrak{R}^+\right)_{+1}\oplus\left(\mathbb{C}e_{\theta}\right)_{+2}~.
\end{equation}
The dimension of $\mathbb{O}_{\min}(\gf)$ can be deduced from \eqref{eq:five_grade_decomposition} by counting the number of generators that act non-trivially on $e_{\theta}$. These consist of all the elements in $\mathfrak{R}^-$, along with $f_{\theta}$ and $h_{\theta}$. It follows that  
\begin{equation}
\dim_{\mathbb{C}}\,\mathbb{O}_{\min} = \dim\mathfrak{R}+2 = 2(h^{\vee}-1)~.
\end{equation}
In the second equality we have used the identity $\dim\mathfrak{R}=2(h^{\vee}-2)$ where $h^{\vee}$ is the dual Coxeter number of $\gf$, which can be verified from Table~\ref{Table_Deligne}.

The (closure of the) minimal nilpotent orbit for any $\gf$ is an affine algebraic variety with quadratic defining relations, known as the \emph{Joseph relations}. In the symmetric algebra $S(\gf^\ast)$ (\ie, polynomials on $\gf$) one defines the ideal $\II_2$ according to
\begin{equation}\label{Introducing_I2}
\text{Sym}^2{\bf Adj}= [2{\bf Adj}]\oplus \mathcal{I}_2~,
\end{equation}
where ${\bf Adj}$ denotes the adjoint representation of $\gf$, $[2{\bf Adj}]$ is the representation with Dynkin labels twice those of the adjoint, and we identify ${\rm Sym}^2{\bf Adj}$ with the space of degree-two polynomials on $\gf$. The closure of the minimal nilpotent orbit is then given by
\begin{equation}
\overline{\mathbb{O}_{\min}(\gf)}=\Big\{x\in\gf~\Big|~{\mathcal{I}_2}=0\Big\}~.
\end{equation}

In the case $\gf=\slf(2)$ studied above, we were able to (rather trivially) solve the defining relations of the minimal nilpotent orbit given in \eqref{xxiszeroSL2} in the open patch where $\mathsf{e}\neq 0$. We want to perform an analogous construction for the general case. First we fix a basis for $\gf$ which reflects the decomposition \eqref{d_decomp_Deligne}. Let $\mathsf{e}_A,~\mathsf{f}_A,~A=1,\dots,\dim(\mathfrak{R})$, be bases for $\mathfrak{R}^+$ and $\mathfrak{R}^-$, let $\mathsf{J}_{\alpha}^{\natural}$ be a basis for $\gf^{\natural}$, and let $\mathsf{e}_{\theta},\mathsf{h}_{\theta},\mathsf{f}_{\theta}$ be a basis for $\slf(2)_{\theta}$. In this basis, the $\gf-$Lie algebra/Poisson brackets of moment maps is given by
\begin{equation}
\label{PoissonAllDelign}
\begin{alignedat}{6}
&\big{\{} \mathsf{e}_{A},\mathsf{e}_{B} \big{\}}&~=~& +\Omega_{AB}\,\mathsf{e}_{\theta}~,\qquad&&
\big{\{} \mathsf{e}_{A},\mathsf{e}_{\theta} \big{\}} &~=~& 0~,\qquad&&
\qquad\big{\{} \mathsf{e}_{\theta},\mathsf{e}_{\theta} \big{\}} &~=~& 0~,\\
&\big{\{} \mathsf{f}_{A},\,\mathsf{f}_{B}\big{\}} &~=~& -\Omega_{AB}\,\mathsf{f}_{\theta}~,\qquad&&
\big{\{} \mathsf{f}_{A}\,,\,\mathsf{f}_{\theta}\big{\}} &~=~& 0~,\qquad&&
\qquad\big{\{} \,\mathsf{f}_{\theta},\,\mathsf{f}_{\theta}\big{\}} &~=~& 0~,\\
&\big{\{} \mathsf{f}_{\theta},\,\mathsf{e}_{A}\big{\}} &~=~& \mathsf{f}_A~,\qquad&&
\big{\{} \mathsf{e}_{\theta},\,\mathsf{f}_{A}\big{\}} &~=~& \mathsf{e}_A~,&&&&\\
&\big{\{}\mathsf{h}_{\theta},\,\mathsf{e}_{A}\big{\}} &~=~& \mathsf{e}_A~,\qquad&&
\big{\{}\mathsf{h}_{\theta},\, \mathsf{f}_{A}\big{\}} &~=~& -\mathsf{f}_A~,&&&&\\
&\big{\{} \mathsf{f}_{A},\,\mathsf{e}_{B}\big{\}} &~=~& I_{AB}^{\alpha}\,\mathsf{J}_{\alpha}^{\natural}- \tfrac{1}{2}\,\Omega_{AB}\, \mathsf{h}_{\theta}~,\qquad&&
\big{\{} \mathsf{J}_{\alpha}^{\natural},\,\mathsf{J}_{\beta}^{\natural}\big{\}} &~=~& (f^{\natural})_{\alpha \beta}^{\,\,\,\,\,\,\gamma}\,\mathsf{J}_{\gamma}^{\natural}~,&&&&\\
&\big{\{} \mathsf{J}_{\alpha}^{\natural},\,\mathsf{e}_A\big{\}} &~=~& \left(t_{\alpha}\right)_{A}^{\,\,\,\,B}\,\mathsf{e}_{B}~,\qquad&&
\big{\{} \mathsf{J}_{\alpha}^{\natural},\,\mathsf{f}_A\big{\}} &~=~& \left(t_{\alpha}\right)_{A}^{\,\,\,\,B}\,\mathsf{f}_{B}~,&&&&\\
&\big{\{} \mathsf{J}_{\alpha}^{\natural},\,\mathsf{e}_{\theta}\big{\}} &~=~& 0~,\qquad&&
\big{\{} \mathsf{J}_{\alpha}^{\natural},\,\mathsf{h}_{\theta}\big{\}} &~=~& 0~,\qquad&&\
\qquad\big{\{} \mathsf{J}_{\alpha}^{\natural},\,\mathsf{f}_{\theta}\big{\}} &~=~& 0~.
\end{alignedat}
\end{equation}
Here $\Omega_{AB}$ is a non-degenerate symplectic form on $\mathfrak{R}$, the symmetric tensor $I_{AB}^{\alpha}=I_{BA}^{\alpha}$ intertwines $\text{Sym}^2 \mathfrak{R}$ with the adjoint representation of $\gf^{\natural}$, $(f^{\natural})_{\alpha \beta}^{\phantom{\alpha\beta}\gamma}$ are the structure constants of $\gf^{\natural}$, and the tensor $\left(t_{\alpha}\right)_{A}^{\phantom{A}B}$ specifies the embedding of $\gf^{\natural}$ in $\mathfrak{sp}(\mathfrak{R})$ as in Table~\ref{Table_Deligne}. This embedding can be equivalently specified by a tensor $T_{\alpha}^{AB}$ such that $\mathsf{J}^{\natural}_{\alpha}=T_{\alpha}^{AB}\,\mathsf{M}_{AB}$ where $\mathsf{M}_{AB}$ are the generators of $\mathfrak{sp}(\mathfrak{R})$ and $\left(t_{\alpha}\right)_{A}^{\phantom{A}B}=-2\,\Omega_{AC}\,T_{\alpha}^{CB}$.

The Joseph relations can be decomposed according to their weight under the action of $\mathsf{h}_{\theta}$. The relations of maximal weight (two) are given by
\begin{equation}
\label{Yrelation}
\mathsf{ Y}_{\alpha}\,\colonequals\,T_{\alpha}^{AB}\,\mathsf{e}_A\,\mathsf{e}_B-\mathsf{e}_{\theta}\,\mathsf{J}^{\natural}_{\alpha}\,=\,0~.
\end{equation}
It is easy to verify that $\mathsf{Y}_{\alpha}$ obeys $\{\mathsf{e}_A,\mathsf{Y}_{\alpha}\}=0$, which implies that the relations $\mathsf{ Y}_{\alpha}$ have maximal weight in their $\gf-$representation. The next set of Joseph relations are the $\gf$ descendants of $\mathsf{Y}_{\alpha}$, which are given by
\begin{equation}\label{Yprelation}
\{\mathsf{f}_A,\mathsf{Y}_{\alpha}\} = \left(t_{\alpha}\right)_{A}^{\phantom{A}B}\,\mathsf{Y}^\prime_{B}~,\qquad
\mathsf{Y}^\prime_{A}\,\colonequals\,\mathsf{e}_{\theta}\,\mathsf{f}_{A}+\tfrac{1}{2}\,\mathsf{h}_{\theta}\,\mathsf{e}_A+\left(K^{\alpha}\right)_{A}^{\phantom{A}B} \mathsf{J}^{\natural}_{\alpha}\,\mathsf{e}_B\,=\,0~,
\end{equation}
where the explicit form of the tensor $\left(K^{\alpha}\right)_{A}^{\phantom{A}B}$ will not be used.\footnote{This tensor is determined from the identity $\delta^A_{B}\,\delta_{\alpha}^{\beta}+2\,T_{\alpha}^{CB}\,I^{\beta}_{AC}\,=\,\delta_{\alpha}^{\beta}\,\left(t_{\gamma}\right)_{A}^{\phantom{A}C}\left(K^{\gamma}\right)_{C}^{\phantom{C}B}$.} With $e_\theta$ inverted, equations~\eqref{Yrelation} and \eqref{Yprelation} can be solved for $\mathsf{J}^{\natural}_{\alpha}$ and $\mathsf{f}_{A}$ in terms of the generators $\mathsf{e}_{\theta}^{\pm 1}$, $\mathsf{e}_{A}$, and $\mathsf{h}_{\theta}$. 

The generator $\mathsf{f}_{\theta}$ could then be determined in a similar way by looking at $\gf$ descendants of the relations $\mathsf{Y}^\prime_{A}$. However, it is more efficient to obtain this generator using the Poisson bracket
\begin{equation}
\{\mathsf{f}_{A},\mathsf{f}_{B}\}=\Omega_{AB}\mathsf{f}_{\theta}~.
\end{equation}
The left-hand side can be computed using the expression for $\mathsf{f}_{A}$ in terms of $\mathsf{e}_{\theta}^{\pm 1}$, $\mathsf{e}_{A}$ and $\mathsf{h}_{\theta}$, and the Poisson brackets \eqref{PoissonAllDelign}.

The result is that in the dense, open patch $\{\mathsf{e}_{\theta}\neq0\}$, the Joseph relations can be solved completely by expressing $\mathsf{J}^{\natural}_{\alpha}$, $\mathsf{f}_{A}$, $\mathsf{f}_{\theta}$ in terms of the remaining generators:
\begin{equation}\label{eJethetaCLASSICALA}
\mathsf{J}^{\natural}_{\alpha}\,=\,T_{\alpha}^{AB}\,\mathsf{e}^{}_A\,\mathsf{e}^{}_B\,\mathsf{e}_{\theta}^{-1}~,\qquad
\mathsf{f}_{A}\,=\,\left(\widehat{\mathsf{e}}_A^{} \mathsf{e}_{\theta}^{-1}-\tfrac{1}{2}\,\mathsf{h}_{\theta}\,\mathsf{e}_A\right)\mathsf{e}_{\theta}^{-1}~,\qquad
\mathsf{f}_{\theta}=\left(\mathsf{S}-\tfrac{1}{4}\mathsf{h}_{\theta}^2\right)\, \mathsf{e}_{\theta}^{-1}~,
\end{equation}
where we have defined
\begin{equation}\label{eJethetaCLASSICALB}
\widehat{\mathsf{e}}_A^{} \,=\,-\Omega_{AB}\,X^{BCDE}\mathsf{e}_{C}\mathsf{e}_{D}\mathsf{e}_{E}~,\qquad
\mathsf{S}=\tfrac{1}{4}X^{ABCD}\,\mathsf{e}_{A}\mathsf{e}_{B}\mathsf{e}_{C}\mathsf{e}_{D}\,\mathsf{e}_{\theta}^{-2}~.
\end{equation} 
We note that $\mathsf{S}$ is invariant under $\gf^{\natural}$ and the tensor $X$ projects $\text{Sym}^4\,\mathfrak{R}$ onto the singlet representation.\footnote{The structure of $\mathsf{S}$ is discussed in some detail for the DC series in Section \ref{sec:DConebyone}. For the remaining cases, we have
\begin{equation}
\mathsf{e}_{\theta}^{2}\,\mathsf{S}\,\sim\,
\begin{cases}
(\mathsf{h}_{\perp})^2 & \text{for}\,\, \gf=\fa_{r}~,\\
0 & \text{for}\,\, \gf=\mathfrak{c}_{r}~,\\
\epsilon_{ab}^{}\epsilon_{cd}^{}g_{IJ}^{}\,g_{KL}^{}
\mathsf{e}_{I}^{a}
\mathsf{e}_{K}^{b}
\mathsf{e}_{J}^{c}
\mathsf{e}_{L}^{d}
\
& \text{for}\,\, \gf=\mathfrak{so}_{M}~.
\end{cases}
\end{equation}
In the final entry, $a,b\dots\in \{1,2\}$ and $I,J\dots\in\{1,2,\dots,M-4\}$. Refer to Table~\ref{Table_Deligne} to discern notation.} The Poisson brackets \eqref{PoissonAllDelign} hold as a consequence of the following basic brackets amongst independent moment maps,
\begin{equation}\label{PoissondelignBASICs}
\big{\{} \mathsf{e}_{A},\mathsf{e}_{B}\}\,=\,\Omega_{AB}\,\mathsf{e}_{\theta}~,\quad
\big{\{} \mathsf{e}_{A},\mathsf{e}_{\theta}\}\,=\,\big{\{} \mathsf{e}_{\theta},\mathsf{e}_{\theta}\}\,=\, 0~,\quad 
\big{\{}\mathsf{h}_{\theta},\,\mathsf{e}_{A}\big{\}}\,=\, \mathsf{e}_A~,\quad 
\big{\{}\mathsf{h}_{\theta},\,\mathsf{e}_{\theta}\big{\}}\,=\, 2\,\mathsf{e}_{\theta}~.\\
\end{equation}

The expressions we have deduced above generalize \eqref{finsl2}, but are not yet in a form that suggests an obvious affine free field uplift. However, this can be remedied by a slight modification. Consider the cotangent bundle
\begin{equation}
\label{cotbundleforDELIGNE}
T^\ast\left(\Cb^\ast \times \Cb^{(h^{\vee}-2)}\right)~,
\end{equation}
with coordinates $\mathsf{d}_{\theta}$ and $\mathsf{h}_{\theta}$ for the $T^\ast(\Cb^\ast)$ factor and $\upxi_{A}$, $A=1,\dots, 2(h^{\vee}-2)$ for the remaining affine factor. The symplectic structure on this space is the canonical one, given by the following Poisson brackets of the coordinate functions
\begin{equation}\label{PoissondelignforFREE}
\big{\{} \upxi_{A}, \upxi_{B}\}\,=\,\Omega_{AB}\,~,\quad
\big{\{} \upxi_{A},\mathsf{e}_{\theta}\}\,=\,\big{\{} \mathsf{e}_{\theta},\mathsf{e}_{\theta}\}\,=\, 0~,\quad 
\big{\{}\mathsf{h}_{\theta},\,\upxi_{A}\big{\}}\,=\, 0~,\quad 
\big{\{}\mathsf{h}_{\theta},\,\mathsf{d}_{\theta}\big{\}}\,=\, \mathsf{d}_{\theta}~.\\
\end{equation}
We can now embed the Poisson algebra of the $\mathsf{e}_{A}$ and $\mathsf{e}_\theta$ in the Poisson algebra of coordinate functions on $T^\ast(\Cb^\ast)$ according to
\begin{equation}\label{eAfromeupxi}
\mathsf{e}_{A}=\left(\mathsf{d}_{\theta}\right)\,\upxi_{A}~,\qquad \mathsf{e}_{\theta}=\mathsf{d}_{\theta}^2~.
\end{equation}
Alternatively, we have an isomorphism of the relevant open patch of the minimal nilpotent orbit with $T^\ast(\Cb^\ast\times \Cb^{(h^\vee-2)})/\Zb_2$, where the $\Zb_2$ acts by negation on $\upxi_A$ and $\mathsf{d}_\theta$. As $\mathsf{d}_\theta$ is $\Cb^\ast$-valued, this is a free $\Zb_2$ action and the quotient is smooth. It is this smooth $\Zb_2$ quotient that will play the role that was previously played by $T^\ast\Cb^\ast$ in the case $\gf=\slf(2)$.

\medskip

\noindent \emph{Remark:} It is somewhat interesting to note that the computation described in the previous paragraphs is very similar to what Joseph did in his original paper \cite{ASENS_1976_4_9_1_1_0}. Let $U(\gf)$ denote the universal enveloping algebra of $\gf$, $J_0$ the Joseph ideal and $\mathfrak{r}=\mathbb{C}h_{\theta}\oplus\mathfrak{R}^+\oplus \mathbb{C}e_{\theta}$. Theorem 4.3 of \cite{ASENS_1976_4_9_1_1_0} states that for $\gf\neq\slf_n$ there exists a unique embedding of $U(\gf)/J_0$ in $U(\mathfrak{r})_{e_{\theta}}\colonequals \{e_{\theta}^{-s}\,U(\mathfrak{r}), s=0,1,2,\dots\}$ that reduces to the identity when restricted to $\mathfrak{r}$. The classical limit of this embedding coincides with \eqref{eJethetaCLASSICALA}, \eqref{eJethetaCLASSICALB}. For $\gf= \slf_n$ there is a one parameter families of such embeddings.

\subsubsection*{Affinization and free field realizations}
\label{sec:affinization_and_free}

We are now in position to develop an affine uplift of the above construction. We introduce affine currents in the same basis for $\gf$ as above: let $e_A$, $f_A$, $J_{\alpha}^{\natural}$, $A=1,\dots,\dim(\mathfrak{R}), \alpha=1,\dots,\dim(\gf^{\natural})$ be bases for $\mathfrak{R}^+$, $ \mathfrak{R}^-$, and $\gf^{\natural}$, respectively. The OPEs of $e_{\theta},f_{\theta},h_{\theta}$ are as in \eqref{affineSL2}. We collect here most of the remaining OPEs:
\begin{subequations}
\label{OPEsAllDelign}
\begin{alignat}{6}
\label{positivegOPE}
& e_{A}(z_1)e_{B}(z_2) &\,\sim\,& +\frac{\Omega_{AB}}{z_{12}}\,e_{\theta}(z_2)~,
\qquad&&
e_{A}(z_1)e_{\theta}(z_2) &&\,\sim\, 0~,
\qquad&&
e_{\theta}(z_1)e_{\theta}(z_2) &\,\sim\,& 0~,\\
\label{negativegOPE}
& f_{A}(z_1)f_{B}(z_2) &\,\sim\,& -\frac{\Omega_{AB}}{z_{12}}\,f_{\theta}(z_2)~,
\qquad&&
f_{A}(z_1)f_{\theta}(z_2) &&\,\sim\, 0 ~,
\qquad&&
f_{\theta}(z_1)f_{\theta}(z_2) &\,\sim\,& 0~,\\
\label{fe_efOPEmore}
& f_{\theta}(z_1)e_{A}(z_2) &\,\sim\,& +\frac{1}{z_{12}}\,f_A(z_2)~,
\qquad&&
e_{\theta}(z_1)f_{A}(z_2) &&\,\sim\, +\frac{1}{z_{12}}\,e_A(z_2)~,
&&&&
\\
\label{fe_efOPEmore2}
& h_{\theta}(z_1)e_{A}(z_2) &\,\sim\,& +\frac{1}{z_{12}}\,e_A(z_2)~,
\qquad&&
h_{\theta}(z_1)f_{A}(z_2) &&\,\sim\, -\frac{1}{z_{12}}\,f_A(z_2)~,
&&&&\\
\label{positivewithnegativegOPE}
&f_{A}(z_1)e_{B}(z_2) &\,\sim\,& ~\frac{k}{z_{12}^2}\Omega_{AB}\,+ \frac{1}{z_{12}}\big(I_{AB}^{\alpha}\,J_{\alpha}^{\natural}(z_2&&) - \tfrac{1}{2}\,\Omega_{AB}\, h_{\theta}&&(z_2) \big)~.
\end{alignat}
\end{subequations}
The notation is further explained below \eqref{PoissonAllDelign}. We have omitted the OPEs of the form $J^{\natural}\times J^{\natural}$, $J^{\natural}\times e$, and $J^{\natural}\times f$, all of which follow from the Poisson brackets \eqref{PoissonAllDelign}.

\paragraph{The free field construction.} As in the $\fa_1$ case, the free fields we will use in our construction are in correspondence with the coordinates of the cotangent bundle appearing in \eqref{cotbundleforDELIGNE} and \eqref{PoissondelignforFREE}. In this case, in addition to the chiral bosons $\delta$ and $\varphi$ satisfying \eqref{deltaphiOPE} and \eqref{htheta}, we have additional symplectic bosons $\{\xi_A(z)\}$ with OPEs
\begin{equation}\label{xixiOPE}
\xi_{A}(z_1)\xi_{B}(z_2)\,\sim\, \frac{\Omega_{AB}}{z_{12}}~.
\end{equation} 
This is just $(h^{\vee}-2)$ copies of the $(\beta,\gamma)-$VOA, and we will denoted it by $\mathbb{V}_{\xi}$. The $(\beta,\gamma)$ VOA comes with a natural filtration defined by assigning the weight $R[\beta]=R[\gamma]=\tfrac{1}{2}$ and letting derivatives carry zero degree. In other words, we have
\begin{equation}
F_R V_{\beta\gamma} = {\rm span}\left\{(\partial^{n_1}\beta\cdots\partial^{n_k}\beta\partial^{m_1}\gamma\cdots\partial^{m_\ell}\gamma)~,\quad k+\ell\leqslant 2R\right\}~.
\end{equation}
The associated graded of this filtration is simply the arc space $J_\infty\Cb^2=J_\infty(T^\ast\Cb)$, with the $R$-grading induced by the $\Cb^\ast$ scaling action of $\Cb^2$. Because of the $\Zb_2$ quotient described above in our finite-dimensional model for the Higgs branch, our free field realizations will now be given in terms of not the lattice subVOA $\Pi$ but rather the related subVOA
\begin{equation}\label{PihalfDEF}
\Pi_{\frac{1}{2}}\,\colonequals \,\bigoplus_{n=-\infty}^{\infty}\,\left(V_{\partial\varphi}\otimes V_{\partial\delta}\right)\,e^{\frac{n}{2}(\delta+\varphi)}~.
\end{equation}
In the remainder of this section, we will give a realization of the (simple) affine VOA $V_{-\frac{h^{\vee}}{6}-1}(\mathfrak{g})$, where $\mathfrak{g}\,\in\,\{\fa_2, \fg_2, \fd_4, \ff_4, \fe_6, \fe_7, \fe_8\}$ as a subalgebra of
\begin{equation}
\mathbf{V}_{\text{FF}}\colonequals  \mathbb{V}_{\xi}\otimes \Pi_{\frac{1}{2}}~,\qquad
\mathbb{V}_{\xi} \simeq \left(V_{\beta \gamma}\right)^{\otimes (h^{\vee}-2)}~.
\end{equation}

The form of the generators $e(z)$ and $e_A(z)$ is fixed by charge and filtration considerations to be
\begin{equation}\label{eandeDeligne}
e_{\theta}(z)=1\otimes e^{\delta+\varphi}~,\qquad 
e_{A}(z)=\xi_A\,\otimes\,e^{\frac{\delta+\varphi}{2}}~.
\end{equation}
These expression should be compared to the classical counterparts \eqref{eAfromeupxi}. It is clear that these generators satisfy the correct OPEs \eqref{positivegOPE}. Let us then turn our attention to the $\gf^{\natural}$ factor. The latter is a subalgebra of $\mathfrak{sp}(\mathfrak{R})$. The realization of the corresponding affine VOA then follows from the canonical realization
of the symplectic VOA as a subalgebra of $\mathbb{V}_{\xi}$, so that  
\begin{equation}\label{gnatural}
J^{\natural}_{\alpha}(z)\colonequals T_{\alpha}^{AB}\,\xi_A\xi_B \otimes 1~,\qquad
V^{k^{\natural}}(\gf^\natural)\subset V_{-\frac{1}{2}}(\mathfrak{sp}(\mathfrak{R}))\subset\mathbb{V}_{\xi}~.
\end{equation}
where the tensor $T_{\alpha}^{AB}$ specifies the embedding. This implies that, for $\gf\neq\fa_1,\fa_2$, the level $k$ can be determined as follows.\footnote{This restriction comes from the requirement that for our argument to apply, $\gf^{\natural}$ must be non-abelian.} The level $k^{\natural}$ is fixed by the embedding of $\gf^{\natural}\hookrightarrow\mathfrak{sp}(\mathfrak{R})$ to take the value $k^{\natural}=-\tfrac{1}{2}I_{\gf^{\natural} \hookrightarrow \mathfrak{sp}(\mathfrak{R})}$. This embedding index can be easily determined by looking at the index of $\mathfrak{R}$ as a representation of $\gf^{\natural}$. Similarly, the level $k$ is obtained from $k^{\natural}$ by the embedding of $\gf^{\natural}\hookrightarrow\gf$. The result is that the level $k$ is determined to be
\begin{equation}
\label{LevelkfromtheSPembedding}
 k=-\frac{1}{2}\frac{I_{\gf^{\natural} \hookrightarrow \mathfrak{sp}(\mathfrak{R})}}{I_{\gf^{\natural}\hookrightarrow\gf}}\qquad \text{for $\gf \neq \fa_1,\fa_2$}~.
\end{equation}
For the Deligne-Cvitanovi\'c exceptional series this reproduces the expected value $k=-\tfrac{1}{6}h^{\vee}-1$, see Table~\ref{Table_Deligne} for more details.

We now address the realization of $\slf(2)_{\theta}$. Notice that the embedding index of $\slf(2)_{\theta}=\langle e_{\theta},f_{\theta},h_{\theta} \rangle$ into $\gf$ is $1$, so that the level of $\slf(2)_{\theta}$ coincides with the level of $\gf$ denoted by $k$. The generators $e_{\theta}(z)$ and $h_{\theta}(z)$ are given in \eqref{eandeDeligne} and \eqref{htheta}. The remaining generator $f_{\theta}(z)$ is fixed by $f_\theta\times e_\theta$ and $f_\theta \times h_\theta$ OPEs to take a form that generalizes the $\slf(2)$ result \eqref{fSL2},
\begin{equation}\label{fthetaDELIGNE}
f_{\theta}(z)= \Big(S^{\natural}\otimes 1-1\otimes\left((\tfrac{k}{2}\,\partial\delta )^2-\tfrac{k(k+1)}{2}\partial^2 \delta\right)\Big)
\Big(1\otimes e^{-(\delta+\varphi)}\Big)~,
\end{equation}
where we recall that $\langle \varphi,\varphi\rangle=-\langle\delta,\delta\rangle=\tfrac{2}{k}$, and $S^{\natural}\in \mathbb{V}_{\xi}$ must have $h=2$ and be a $\gf^\natural$ singlet. Furthermore, since the $\slf(2)_\theta$ and $\gf^{\natural}$ current algebras must commute, $S^\natural$ should commute with the $J^\natural$ currents, and finally the $S^\natural \times S^\natural$ OPE must conspire with the value of $k$ to render the $f_\theta \times f_\theta$ OPE regular.

For the DC series, such an $S^\natural$ can indeed be found. It is useful to distinguish two cases,
\begin{equation}\label{SnaturalPRESCRIPTION}
S^{\natural} = \begin{cases}
(k+2)T^{\natural}~, \qquad T^{\natural}= -T_{\text{Sug}}[{\gf}^{\natural}] \,+\, \partial\xi\,\Omega^{-1}\xi~, & k\neq -2~,\\[10pt]
-\frac{1}{2}\Big((J^{\natural}_1,J^{\natural}_1)_{A_1}+(J^{\natural}_2,J^{\natural}_2)_{A_1}+(J^{\natural}_3,J^{\natural}_3)_{A_1}\Big)~, &\text{$k=-2$, i.e.~$\gf=\fd_4$}~,
\end{cases}
\end{equation}
where $T_{\text{Sug}}[{\gf}^{\natural}]$ is the Sugawara stress-tensor constructed using the $g^{\natural}$ affine currents. In the special case of $\gf=\fd_4$, the three bilinears in the expression above are actually identical $(J^{\natural}_1,J^{\natural}_1)=(J^{\natural}_2,J^{\natural}_2)=(J^{\natural}_3,J^{\natural}_3)$, as is familiar from class $\SS$ constructions involving full punctures. It should be noted that if $k\neq -2$, then $T^{\natural}$ satisfies the Virasoro algebra with
\begin{equation}\label{constrcnatural}
c^{\natural}=1-6\frac{\,(k+1)^2}{k+2}~.
\end{equation}
This non obvious fact can be verified by direct calculation. The value of the central charge given in \eqref{constrcnatural} coincides with that of the Virasoro algebra obtained by quantum Drinfel'd-Sokolov reduction of $\slf_\theta(2)$. In the remaining case with $k=-2$, $S^{\natural}$ has regular OPE with itself. These properties of $S^{\natural}$ guarantee that $f_{\theta}(z_1)f_{\theta}(z_2)\sim 0$. It is also essential that $T^{\natural}$ has regular OPE with the generators of $V^{k^{\natural}}(\gf^\natural)$, which guarantees that indeed $J^{\natural}(z_1) f_{\theta}(z_2)\sim 0$.

The generators that remain to be constructed are the $f_A(z)$. These are $\slf_\theta(z)$ descendants of $e_{A}(z)$, so they can be obtained by the first OPE in \eqref{fe_efOPEmore}. This yields the expression
\begin{equation}\label{fADEL}
f_{A}(z)=\Big(\widehat{\xi}_A\otimes 1+\tfrac{k}{2}\,\xi_A\otimes \partial\delta\Big)
\Big(1\otimes e^{-\frac{1}{2}(\delta+\varphi)}\Big)~,
\end{equation}
where $\widehat{\xi}$ is defined as
\begin{equation}\label{SxiOPE}
S^{\natural}(z_1)\,\,\xi_A(z_2)\,\sim\,\frac{h^{\natural}_\xi}{z_{12}^2}\,\,\xi_A(z_2)+\frac{1}{z_{12}}\,\widehat{\xi}_A(z_2)~,\qquad h^{\natural}_{\xi}=-\tfrac{1}{4}(2k+1)~.
\end{equation}
When computing $f_{A}$ via the first OPE in \eqref{fe_efOPEmore}, there is a potential second-order pole. This vanishes thanks to a cancellation between the term proportional to $h^{\natural}_{\xi}$ in \eqref{SxiOPE} and a second order pole coming from the part of the OPE arising entirely from $\Pi_{\frac{1}{2}}$. The complete set of generators have thus been constructed, see \eqref{eandeDeligne}, \eqref{gnatural}, \eqref{htheta}, \eqref{fthetaDELIGNE}, and \eqref{fADEL}. It is not entirely obviously that the remaining OPEs will necessarily take the correct form, but these can be verified by explicit calculation.

The expressions for the generators $f_\theta(z)$ and $f_A(z)$ given in \eqref{fthetaDELIGNE} and \eqref{fADEL} should be compared to their classical counterparts \eqref{eJethetaCLASSICALA}, \eqref{eJethetaCLASSICALB}, and \eqref{eAfromeupxi}. In fact, the former can be obtained starting from the classical expressions and adding (finitely many)
lower terms in the $R$-filtration so that the OPEs of the affine currents are correctly reproduced.

\paragraph{Nilpotent Higgsing and Drinfel'd-Sokolov reduction.} Let us pause for a moment to discuss in greater detail the relation of the above construction to the procedure of nilpotent Higgsing. In particular, we observe that in a sense the free field realizations we have presented can be considered as inversions of the quantum Drinfel'd-Sokolov reduction.

In our examples, the Higgs branch is smooth away from the origin and all the points on the smooth locus are equivalent. Let us consider the point corresponding to $\mathsf{e}_{\theta}=1$ with all the remaining generators taken to vanish. The transverse space to the $SL(2)_\theta$ orbit of this point inside the minimal nilpotent orbit of $\mathfrak{g}$ is isomorphic to $T^\ast(\Cb^{(h^{\vee}-2)})$. This space can be identified with the space described by the coordinates $\upxi_{A}$ appearing in \eqref{eAfromeupxi}, and are associated to (Goldstone) bosons of the free theory obtained in the IR. The same subspace $T^\ast(\Cb^{(h^{\vee}-2)})$ can also be identified by first solving the Joseph relations as we did above and further picking the point $\mathsf{e}_{\theta}=1$, $\mathsf{h}_{\theta}=0$ in the $T^\ast\mathbb{C}^\ast$ factor of \eqref{cotbundleforDELIGNE}. In these examples the R-symmetry in the IR, denoted by $SU(2)_{\bar{R}}$, can be identified in the UV. The $\bar{R}$ Cartan generator mixes with the flavor symmetry as
\begin{equation}
\bar{R}=R-\tfrac{1}{2}\,h_{\theta}~.
\end{equation}
This combination of quantum numbers vanishes for the operator that gets an expectation value, namely $\mathsf{e}_{\theta}$, while the coordinates associated to the free bosons satisfy $R[\upxi_A]=\bar{R}[\upxi_A]=\tfrac{1}{2}$.

At the level of the associated VOA, this Higgsing operation corresponds to quantum Drinfel'd-Sokolov (DS) reduction\footnote{This is the case in an array of situations that includes the examples above and many class $\SS$ constructions \cite{Beem:2014rza}.} which is the procedure for imposing the constraint $e_{\theta}(z)=1$ via BRST. It is useful to consider a simple variant of the usual DS reduction, described in \cite{Beem:2014rza}, in which one regards all the affine currents of $\mathfrak{g}$ that are not in $\mathfrak{sl}(2)_{\theta}$ as an $\mathfrak{sl}(2)_{\theta}$ module and performs the reduction as explained in \cite{Beem:2014rza}. For the DC exceptional series this procedure yields precisely the symplectic boson VOA $\mathbb{V}_\xi$.

In the free field realization, this process is equivalent to setting the chiral bosons $\delta(z)=\varphi(z)=0$. Once this is done, the generators $e_A(z)$ (see \eqref{eandeDeligne}) become equivalent to the $\xi_A(z)$, while the remaining generators of the affine VOA reduce to composites of the latter. It appears that this is an instance of a very general construction, with the VOA fo the theory that arises in the IR by Higgsing getting ``dressed'' by chiral bosons to obtain the VOA of the UV theory. We will encounter an interesting variant of this phenomenon in Section \ref{sec_AD}. This seems to be a useful perspective more generally, and will be described in a more illustrative set of examples in \cite{BMRToAppear}.

\paragraph{Central charges and the stress tensor.} Our free field VOA has a canonical stress tensor arising as the sum of stress tensors associated to the various free fields, namely,
\begin{equation}\label{StresstensorTDC}
T=T_{\xi}\,+\,T_{\delta}\,+\,T_{\varphi}\,,
\end{equation}
where $T_{\xi}=\partial\xi\,\Omega^{-1}\xi$, and where we recall that the stress tensor of a chiral boson $\phi$ can be written as
\begin{equation}\label{Tphi}
T_{\phi}=\tfrac{1}{2\langle\phi,\phi\rangle}\,\left((\partial \phi)^2+\alpha\, \partial^2 \phi\right)~,
\end{equation}
where $\alpha$ is the ``background charge''. The generator $h_{\theta} \sim \partial \varphi$ must have conformal weight $h=1$, so $T_{\varphi}$ cannot have a background charge. On the other hand, the operator $e^{\lambda(\delta+\varphi)}$ should have conformal weight $\lambda$, which fixes $\delta$ to have background charge $\alpha_\delta = -2$.\footnote{In our example $\langle \delta,\delta\rangle=-\langle \varphi,\varphi\rangle=-\tfrac{2}{k}$. It follows that our exponentiated null boson $e^{\lambda(\delta+\varphi)}$ has dimension $-\tfrac{1}{2}(\alpha_{\delta}+\alpha_{\varphi})\lambda$ where $\alpha_{\delta}$ and $\alpha_{\varphi}$ are defined as the background charges as in \eqref{Tphi}.} Taken together, this means that the central charges associated to the various factors in \eqref{StresstensorTDC} are given by
\begin{equation}
c_\xi = (2-h^\vee)~,\qquad c_{\varphi} = 1~,\qquad c_{\delta} = 1+6k=-h^{\vee}-5~.
\end{equation}
And summing the contributions to the central charge we verify that
\begin{equation}\label{centralchargematchDELIGNE}
c=c_{\xi}+c_{\delta}+c_{\varphi}=\,-1\times (h^{\vee}-2)+(1-h^{\vee}-6)+1=-2-2h^{\vee}~,
\end{equation}
which is the correct (Sugawara) value. Indeed, one can check directly that in our free field realization, the Sugawara stress tensor is identically equal to the canonical stress tensor given in \eqref{StresstensorTDC} with the necessary background charges.

\paragraph{Null states and Higgs chiral ring relations} A key feature of the free field realizations given above is that they in fact realize the \emph{simple quotient} of the relevant affine Kac-Moody VOA. In other words, null states are identically set to zero. For now we will omit the $\fa_1$ case, which will be further discussed in Section \ref{sec_AD}. For any $\gf$ the Joseph ideal introduced in \eqref{Introducing_I2} can be decomposed as 
\begin{equation}
\mathcal{I}_2\,=\,\mathbb{X}\oplus1~.
\end{equation}
A unique feature of the Deligne-Cvitanovi\'c exceptional series is that $\mathbb{X}$ is an irreducible representation of $\gf\ltimes\text{Out}(\gf)$ where $\text{Out}(\gf)$ denotes the group of outer automorphisms of $\gf$.\footnote{The only case for which the outer automorphism acts non trivially is given by $\fd_4$ for which $\mathbb{X}$ is the sum of three irreducible representations that  are permuted by the triality automorphism of $\fd_4$. In the case of $\fa_1$ the factor $\mathbb{X}$ is trivial.} The maximal ideal of $V^k(\gf)$ for $k=-1-h^{\vee}/6$ and $\gf$ in the DC series, with the exception of $\fa_1$, is generated by singular vectors of conformal dimension two transforming in the representation $\mathbb{X}$.\footnote{Following standard VOA notation, we use the symbol $V^k(\gf)$ to denote the universal affine vertex algebra at level $k$ and $V_k(\gf)$ to indicate its simple quotient.} This fact is well known for the cases $\fa_2$, $\fg_2$, and $\ff_4$, for which the level is admissible\footnote{The case of $\fa_2$ is a so-called boundary admissible levels: $h^{\vee}+k=\tfrac{h^{\vee}}{p}$ with $p\in \mathbb{Z}_{>0}$ and coprime with $h^{\vee}$.} see \cite{Kac:1988qc}, and was proven in for the remaining cases in \cite{Arakawa:2015jya}. We then need to show that the free field realizations presented above satisfy
\begin{equation}
\NO{J}{J}\big{|}_{\mathbb{X}}=0~,
\end{equation}
Since $\mathbb{X}$ is an irreducible representation we can focus on any vectors in $\mathbb{X}$. We will choose to consider the vectors with maximal $h_{\theta}$ weight, which are given by
\begin{equation}
T_{\alpha}^{AB}\,e_A\,e_B- e_{\theta}\,J^{\natural}_{\alpha}~,
\end{equation} 
where the tensor $T_{\alpha}^{AB}$ was introduced in \eqref{gnatural} and we have dropped the $\text{NO}$ from the notation. This is the VOA uplift of the null relation \eqref{Yrelation}. It is straightforward to verify that these operators are identically zero in our realizations by using the expressions for $e_{\theta}$, $e_A$, and $J^{\natural}_{\alpha}$ from \eqref{eandeDeligne} and \eqref{gnatural},
\begin{equation}
T_{\alpha}^{AB}\,e_A\,e_B- e_{\theta}\,J^{\natural}_{\alpha}=T_{\alpha}^{AB}
\big{(}\xi_A\otimes e^{\frac{\delta+\varphi}{2}}\big{)}
\big{(}\xi_B\otimes e^{\frac{\delta+\varphi}{2}} \big{)}-
\big{(}1\otimes e^{\delta+\varphi}\big{)}
\big{(}T_{\alpha}^{AB}\xi_A\xi_B\otimes 1 \big{)}\,=\,0~.
\end{equation} 

\vspace{0.5cm}

\noindent \emph{Remark:} It is gratifying to observe that the construction presented in this section works, without adding extra degrees of freedom, only in the cases for which $\mathcal{W}_k(\gf,f_{\theta})\simeq \mathbb{C}$, where $\mathcal{W}_k(\gf,f_{\theta})$ denotes the (simple quotient of) the $\WW$-algebra associated with $(\mathfrak{g},f_{\theta})$ at level $k$. Theorem 7.2 of \cite{Arakawa:2015jya} classifies the pairs $(\gf,k)$ for which this is the case:
\begin{enumerate}
	\item $\gf$ is of type $\fa_1$, $k\,\in\,\{-2,-\frac{1}{2}\}$.
	\item $\gf$ is in the DC exceptional series at level $k=-h^{\vee}/6-1$.
	\item $\gf$ is of type $\mathfrak{c}_{\ell}$  at level  $k=-\frac{1}{2}$.
\end{enumerate}
Entries (1) and (2) have been realized above while (3) corresponds to the standard realization of $V_{-\frac{1}{2}}(\mathfrak{sp}(2r))$ in terms of symplectic bosons. The cases $\fa_{r>2}$ from Table~\ref{Table_Deligne} admit free field realizations similar to the ones presented here, but with an additional affine ${\mathfrak{gl}}_1$ current. This extra free field has a natural geometric interpretation in terms of the associated variety of $V_{-1}(\slf_{r+1})$. The latter was determined in \cite{arakawa2017sheets} and is not symplectic. The case of $\gf=\mathfrak{so}_N$ with $N>6$, $N\neq 8$ requires further additional degrees of freedom.
 
\subsection{Deligne-Cvitanovi\'c constructions in detail}
\label{sec:DConebyone}

For the sake of concreteness, we present details relating to the free field construction for each of the Deligne-Cvitanovi\'c exceptional VOAs below. Many of the constructions given here closely mirror constructions in the case of finite-dimensional Lie algebras appearing in \cite{Gunaydin:2004md}. The relations are due to the similarities between our construction and those of minimal realizations of simple Lie algebras mentioned in the remark below equation \eqref{PoissondelignBASICs}.

\smallskip

\paragraph{The $\fa_2$ entry.} In this case there is a single $(\beta,\gamma)$ symplectic boson pair, and the ingredients in the free field construction are given by
\begin{equation}
\xi=\begin{pmatrix}\beta\\ \gamma\end{pmatrix}~,\qquad
J^{\natural}=\beta\gamma~,\qquad
T^{\natural}=\tfrac{1}{2}J^{\natural}J^{\natural}+\tfrac{1}{2}\left(\beta\partial\gamma-\partial \beta\gamma \right)~,\qquad \widehat{\xi}=
\begin{pmatrix}
-\tfrac{1}{2}\,\beta^2\gamma+\partial\beta\\ 
+\tfrac{1}{2}\,\gamma^2\beta+\partial\gamma
\end{pmatrix}~.
\end{equation}
Here the symplectic form on the $(\beta,\gamma)$ system and the (single) symmetric intertwining tensor $I$ are given by  
\begin{equation}
\Omega=\begin{pmatrix}0 & 1\\-1 & 0\end{pmatrix}~,\qquad I=-\tfrac{3}{2}\begin{pmatrix}0 & 1\\1 & 0\end{pmatrix}~.
\end{equation}

\paragraph{The $\fg_2$ entry.} This is an affine version of a construction due to Joseph \cite{Joseph:1974hr}. In this case there are two symplectic boson pairs, which we denote by 
$\left(\xi_A\right)=(\beta_1,\beta_2,\gamma_1,\gamma_2)$ whose non-vanishing OPEs are given by
\begin{equation}
\beta_a(z)\gamma_b(w)\sim\frac{\delta_{ab}}{z-w}~.
\end{equation}
In this case $\fg^\natural = \fa_1$ and the corresponding affine generators take the form
\begin{equation}
J^{\natural}_+=\beta_2\gamma_1+\tfrac{1}{\sqrt{3}}\,\gamma_2\gamma_2~,\qquad
J^{\natural}_0=3\,\beta_1\gamma_1+\beta_2\gamma_2~,\qquad
J^{\natural}_-=3\left(\beta_1\gamma_2-\tfrac{1}{\sqrt{3}}\,\beta_2\beta_2\right)~,
\end{equation}
while the stress tensor appearing in \eqref{SnaturalPRESCRIPTION} is given by
\begin{equation}
T^{\natural}= -T_{\text{Sug}}[\slf(2)^{\natural}_{k=-5}] +\tfrac{1}{2}\sum_{a=1}^2\left(\beta_a\partial\gamma_a-\partial \beta_a\gamma_a \right)~.
\end{equation}
It is not hard to verify that the symplectic bosons $\left(\xi_A\right)$ transform in the spin-$3/2$ representation of $\fg^\natural = \fa_1$.

\paragraph{The $\fd_4$ entry.} In this case we have $\gf^{\natural}=\fa_1\oplus \fa_1\oplus \fa_1$ and the symplectic bosons $\xi^{a_1a_2a_3}$ transform in the tri-fundamental representation. These can be thought of as the symplectic bosons associated to the $\fa_1$ trinion theory of class $\SS$, where nilpotent Higgsing amounts to closing a single puncture in the four-times punctured sphere \cite{Beem:2014rza}. Their OPE takes the form
\begin{equation}
\xi^{a_1 a_2 a_3}(z)\xi^{b_1 b_2b_3}(w)\,\sim\,\frac{\epsilon^{a_1b_1}\epsilon^{a_2b_2}\epsilon^{a_3b_3}}{z-w}~,
\end{equation}
where $a_{i},b_{i}\,\in\,\{1,2\}$. The affine current subVOA associated to $\widehat{\gf}^{\natural}$ is realized by 
\begin{equation}
\begin{split}
J_1^{(a_1 b_1)}&=\tfrac{1}{2}\,\epsilon_{a_2b_2}\epsilon_{a_3b_3}\,\xi^{a_1 a_2 a_3}\xi^{b_1 b_2b_3}~,\\
J_2^{(a_2 b_2)}&=\tfrac{1}{2}\,\epsilon_{a_1b_1}\epsilon_{a_3b_3}\,\xi^{a_1 a_2 a_3}\xi^{b_1 b_2b_3}~,\\
J_3^{(a_3 b_3)}&=\tfrac{1}{2}\,\epsilon_{a_1b_1}\epsilon_{a_2b_2}\,\xi^{a_1 a_2 a_3}\xi^{b_1 b_2b_3}~.
\end{split}
\end{equation}
The prescription of \eqref{SnaturalPRESCRIPTION} now gives
\begin{equation}
S^{\natural}(z)\colonequals J_1^2(z)=J_2^2(z)=J_3^2(z)~,\qquad
J_A^2\colonequals \epsilon_{ac}\epsilon_{bd} J_A^{ab}J_A^{cd}~.
\end{equation}
The operator $S^{\natural}(z)$ has regular self-OPE and also has regular OPE with the $J_A^{ab}$ ($A=1,2,3$). However it is not a null operator, as is reflected in \eqref{SxiOPE}.

\paragraph{The $\ff_4$ entry.} Here we have $\gf^{\natural}=\mathfrak{c}_3=\mathfrak{sp}(6) \supset \slf(3)$. We adopt a notation in which only the $\slf(3)$ symmetry is manifest. The symplectic bosons transform in the $\mathbf{14}'$ of $\mathfrak{sp}(6)$ which decomposes as
\begin{equation}
\xi = \left(\upbeta^{(AB)},\upgamma_{(AB)}, \beta,\gamma\right)~,
\end{equation}
where $A,B\in \{1,2,3\}$ are $ \slf(3)$ indices and parentheses denote symmetrization. The non-vanishing OPEs among the symplectic bosons are given by
\begin{equation}
\upbeta^{(AB)}(z)\,\upgamma_{(CD)}(w)\sim\frac{\delta^A_C\delta^B_D+\delta^A_D\delta^B_C}{z-w}~,\qquad 
\beta(z)\gamma(w)\sim\frac{1}{z-w}~.
\end{equation}
The generators of $\widehat{\gf}^{\natural}$ take the form
\begin{equation}
\big(J^{\natural}\big)^A_B=\upbeta^{AC}\upgamma_{CB}-\tfrac{1}{3} \delta^A_B\,\upbeta^{CD}\upgamma_{CD}~,\qquad
J^{\natural}=3\,\beta\gamma-\tfrac{1}{2}\,\upbeta^{CD}\upgamma_{CD}~,
\end{equation}
\begin{equation}
\big(J^{\natural}\big)^{(AB)}=\upbeta^{AB}\beta+\tfrac{1}{2\sqrt{2}}\,\epsilon^{ACD}\epsilon^{BEF}\upgamma_{CE}\upgamma_{DF}~,\quad
\big(J^{\natural}\big)_{(AB)}=\upgamma_{AB}\gamma+\tfrac{1}{2\sqrt{2}}\,\epsilon_{ACD}\epsilon_{BEF}\upbeta^{CE}\upbeta^{DF}~.
\end{equation}
In the last equation we have decomposed the $\gf^{\natural}=\mathfrak{sp}(6)$ generators with respect to the $\slf(3)$ subalgebra that acts linearly on the indices $A,B,C,D,\dots$. We will do the same for the remaining entries of the DC series.
\paragraph{The $\fe_6$ entry.} In this case $\gf^{\natural}=\fa_5=\slf(6) \supset \slf(3)\oplus \slf(3)$. We adopt a notation in which only the $\slf(3)\oplus \slf(3)$ symmetry is manifest. The symplectic bosons transform in the $\mathbf{20}$ of $\slf(6)$ which decomposes as
\begin{equation}
\xi\,=\,\left( \upbeta^{A\dot{B}},\upgamma_{\dot{A}B}, \beta,\gamma\right)~,
\end{equation}
where $A,B\in \{1,2,3\}$  and $\dot{A},\dot{B}\in \{\dot{1},\dot{2},\dot{3}\}$. The non-vanishing OPEs among the symplectic bosons are given by
\begin{equation}
\upbeta^{A\dot{A}}(z)\,\upgamma_{B\dot{B}}(w)\sim\frac{\delta^A_B\,\delta^{\dot{A}}_{\dot{B}}}{z-w}~,\qquad 
\beta(z)\gamma(w)\sim\frac{1}{z-w}~.
\end{equation}
The generators of the $\widehat{\gf}^{\natural}$ affine current subVOA take the form
\begin{equation}
\big(J^{\natural}\big)^A_B=\upbeta^{A\dot{A}}\upgamma_{\dot{A}B}-\tfrac{1}{3} \delta^A_B\,\mathcal{H}~,\quad\,\,\,\,
\big(J^{\natural}\big)^{\dot{A}}_{\dot{B}}=-\upgamma_{\dot{B}A}\upbeta^{A\dot{A}}+\tfrac{1}{3} \delta^{\dot{A}}_{\dot{B}}\,\mathcal{H}~,\quad\,\,\,\,
J^{\natural}=3\,\beta\gamma-\mathcal{H}~,
\end{equation}
\begin{equation}
\big(J^{\natural}\big)^{A\dot{B}}=\upbeta^{A\dot{B}}\beta+\tfrac{1}{2}\,\epsilon^{ACD}\epsilon^{\dot{B}\dot{C}\dot{D}}\upgamma_{\dot{C}C}\upgamma_{\dot{D}D}~,\quad
\big(J^{\natural}\big)_{\dot{A}B}=-\upgamma_{\dot{A}B}\gamma-\tfrac{1}{2}\,\epsilon_{\dot{A}\dot{C}\dot{D}}\epsilon^{}_{BCD}\upbeta^{C\dot{C}}\upbeta^{D\dot{D}}~,
\end{equation}
where $\mathcal{H}\colonequals \upbeta^{A\dot{A}}\upgamma_{\dot{A}A}$.

\paragraph{The $\fe_7$ entry.} In this case $\gf^{\natural}=\fd_6=\mathfrak{so}(12) \supset \slf(6)$. We adopt a notation in which only the $\slf(6)$ symmetry is manifest. The symplectic bosons transform in the $\mathbf{32}$ of $\mathfrak{so}(12)$ which decomposes as
\begin{equation}
\xi\,=\,\left( \upbeta^{[AB]},\upgamma_{[AB]}, \beta,\gamma\right)~,
\end{equation}
where $A,B\in \{1,2,3,4,5,6\}$ and square brackets denote antisymmetrization. The non-vanishing OPEs among the symplectic bosons are given by
\begin{equation}
\upbeta^{[AB]}(z)\,\upgamma_{[CD]}(w)\sim\frac{\delta^A_C\delta^B_D-\delta^A_D\delta^B_C}{z-w}~,\qquad\beta(z)\gamma(w)\sim\frac{1}{z-w}~.
\end{equation}
The generators of $\widehat{\gf}^{\natural}$ take the form
\begin{equation}
\big(J^{\natural}\big)^A_B=\upbeta^{AC}\upgamma_{CB}-\tfrac{1}{6} \delta^A_B\,\mathcal{H}~,\qquad\,\,
J^{\natural}=6\,\beta\gamma+\mathcal{H}~,
\end{equation}
\begin{equation}
\big(J^{\natural}\big)^{AB}=\upbeta^{AB}\beta+\tfrac{1}{8}\,\epsilon^{ABCDEF}\upgamma_{CD}\upgamma_{EF}~,\qquad
\big(J^{\natural}\big)_{AB}=\upgamma_{AB}\gamma+\tfrac{1}{8}\,\epsilon_{ABCDEF}\upbeta^{CD}\upbeta^{EF}~,
\end{equation}
where $\mathcal{H}\colonequals \upbeta^{AB}\upgamma_{BA}$.

\paragraph{The $\fe_8$ entry.} In this case $\gf^{\natural}=\fe_7 \supset \slf(8)$, and we use a notation in which only the $\slf(8)$ symmetry is manifest. The symplectic bosons transform in the $\mathbf{56}$ of $\fe_7$ which decomposes according to
\begin{equation}
\xi = \left( \upbeta^{[AB]},\upgamma_{[AB]}\right)~,
\end{equation}
where $A,B\in \{1,\ldots,8\}$ and square brackets denote antisymmetrization. The non-vanishing OPEs among the symplectic bosons are given by
\begin{equation}
\upbeta^{[AB]}(z)\,\upgamma_{[CD]}(w)\sim\frac{\delta^A_C\delta^B_D-\delta^A_D\delta^B_C}{z-w}~.
\end{equation}
The generators of $\widehat{\gf}^{\natural}$ take the form
\begin{equation}
\begin{split}
\big(J^{\natural}\big)^A_B &~=~\upbeta^{AC}\upgamma_{CB}-\tfrac{1}{8} \delta^A_B\mathcal{H}~,\\
\big(J^{\natural}\big)^{ABCD} &~=~\upbeta^{[AB}\upbeta^{CD]}+\tfrac{1}{4!}\,\epsilon^{ABCDEFGH}\upgamma_{EF}\upgamma_{GH}~,
\end{split}
\end{equation}
where $\mathcal{H}\colonequals \upbeta^{AB}\upgamma_{BA}$.

\section{\texorpdfstring{$\mathfrak{su}(2)$}{su(2)} \texorpdfstring{$\mathcal{N}=4$}{N=4} super Yang-Mills}
\label{sec_N4}

We now turn to a case in which free vector multiplets survive in the low energy effective theory on the Higgs branch, namely the simplest interacting $\mathcal{N}=4$ SYM theory, with gauge group $SU(2)$. According to our general recipe, the free field realization of the relevant VOA should involve symplectic fermions in addition to the chiral bosons associated to the geometry.

For this theory, the associated VOA is the (simple quotient of the) small $\mathcal{N}=4$ super-Virasoro algebra at central charge $c=-9$ \cite{Beem:2013sza}. The VOA is generated by $\mathfrak{sl}(2)$ affine currents $e(z),h(z),f(z)$ at level $k=\tfrac{1}{6}c=-\tfrac{3}{2}$, a Virasoro stress-tensor $T(z)$, and four fermionic currents organized into two $\slf(2)$ doublets denoted by $G^{\pm}(z)$ and $\widetilde{G}^{\pm}(z)$. In fact, for this particular level, the Virasoro stress-tensor is not an independent strong generator, but rather it is equal to the Sugawara stress-tensor constructed from the $\slf(2)$ affine currents, as we review below.

When thought of as a special case of an $\mathcal{N}=2$ theory, the Higgs branch is given by
\begin{equation}
\mathcal{M}_{\text{Higgs}}=\mathbb{C}^2/\mathbb{Z}_2~.
\end{equation}
which of course is the same as the Higgs branch in the example of section \ref{subsec:A1_Deligne}. However in the present case, the low energy EFT on the Higgs branch is a $U(1)$ $\mathcal{N}=4$ theory, so the free hypers describing the geometry of the Higgs branch are supplemented by a free vector multiplet. Relatedly, this theory had Hall-Littlewood chiral ring operators that are not simply Higgs branch operators. Indeed, the fermionic currents $G^{\pm}$ and $\widetilde{G}^{\pm}$ are precisely the extra Hall-Littlewood (anti-)chiral ring generators.

As in our previous example, we will have a pair of chiral bosons $\delta,\varphi$ that we will use to describe the same $T^\ast\Cb^\ast$ patch in the Higgs branch as before, but now we also have a symplectic fermion pair, $\{\eta_1(z),\eta_2(z)\}$ with OPE
\begin{equation}
\eta_1(z)\eta_2(w)\sim\frac{2}{(z-w)^2}~,\qquad
\eta_1(z)\eta_1(w)\sim0~,\qquad
\eta_2(z)\eta_2(w)\sim0~.
\end{equation}
We denote by $\mathbb{V}_{\eta}$ the corresponding symplectic fermion VOA. This VOA comes with a natural $R$-filtration with
\begin{equation}
F_R\mathbb{V}_\eta = {\rm span}\left\{\partial^{n_1}\eta_1\cdots\partial^{n_k}\eta_1\partial^{m_1}\eta_2\cdots\partial^{m_\ell}\eta_2~,\quad k+\ell\leqslant R\right\}~.
\end{equation}
Our free field realization will come by identifying the small super-Virasoro VOA as a subVOA of the free fields VOA,
\begin{equation}
\text{Vir}_{\mathcal{N}=4}^{c=-9}\,\subset\mathbb{V}_{\eta}\otimes \Pi_{\frac{1}{2}}~,
\end{equation}
where $\Pi_{\frac{1}{2}}$ is as in \eqref{PihalfDEF}. It is interesting to note the appearance of $\Pi_{\frac12}$ in our construction, even though we have seen before that the (affinization) of single-valued functions on the $T^\ast\Cb^\ast$ in $\mathcal{M}_H$ are described by $\Pi$. This actually has a natural geometric explanation in terms of the Higgs branch effective theory. The free vector multiplet giving rise to the symplectic fermions $\eta_{1,2}$ is associated with unlifted Coulomb branch directions that are fibered over the Higgs branch.\footnote{More generally, one should think of the free fermions appearing in the Higgs branch VOA as spanning a Grassmann-odd vector bundle over the Higgs branch that encodes the embedding of the Higgs branch into the full moduli space of the theory.} Indeed, one can check using the explicit description of the $\mathcal{N}=4$ moduli space as $\Cb^3/\Zb_2$ that the line bundle associated to the unlifted Coulomb directions picks up a factor of $-1$ when encircling the $\Cb^\ast$ direction in the patch of interest, so fermionic variables should be paired with half-integer-powered exponentials in the free field realization.

Based on their quantum numbers and the $R$-filtration, the expression for generators with positive $h(z)$ eigenvalue are fixed to be of the form
\begin{equation}
\label{positiveGeneratorsN4}
e(z)\,=\,1\otimes e^{\delta+\varphi}~,\qquad
\begin{pmatrix}
G^+(z)\\
\widetilde{G}^+(z)
\end{pmatrix}\,=\,
\begin{pmatrix}
\eta_1\\
\,\eta_2
\end{pmatrix}\otimes e^{\frac{\delta+\varphi}{2}}~,
\end{equation}
where we indeed see the predicted pairing of symplectic fermions with half-integer-powered exponentials. These operators are the $\mathfrak{sl}(2)$ highest weight generators of the VOA, and in order to obtain the remaining generators it will be sufficient to construct the $\mathfrak{sl}(2)$ affine Kac-Moody current $f(z)$ and act on the generators in \eqref{positiveGeneratorsN4}. The generator $f(z)$ will take the same form as in the case of DC exceptional series appearing in \eqref{fthetaDELIGNE}, where now we have
\begin{equation}
\label{TetaN=4}
S^{\natural}=(k+2)T^{\natural}=\tfrac{1}{2}\,T^{\natural}~,\qquad T^{\natural}=T_{\eta}:=-\tfrac{1}{2}\eta_1\eta_2~.
\end{equation}
Similarly $h(z)=-\tfrac{3}{2}(1\otimes \partial\varphi)$. By lowering the $\mathfrak{sl}(2)$ spin acting with $f(z)$ on $(G^{+},\widetilde{G}^+)$ we further obtain
\begin{equation}
\begin{pmatrix}G^-(z)\\\widetilde{G}^-(z)\end{pmatrix}\,=\,
\left(\frac{3}{4}\,\begin{pmatrix}\eta_1\\\eta_2\end{pmatrix}\otimes \partial \delta-
\frac{1}{2}\begin{pmatrix}\partial\eta_1\\\partial\eta_2\end{pmatrix}\otimes 1\right)
\Big(1 \otimes\,e^{-\frac{\delta+\varphi}{2}}\Big)~.
\end{equation}
The canonical stress tensor is the sum of three free field contributions, $T=T_{\eta}+T_{\delta}+T_{\varphi}$. In particular, $\eta_1$ and $\eta_2$ have conformal weight $1$ while $e^{\lambda (\delta+\varphi)}$ has weight $\lambda$. As in \eqref{centralchargematchDELIGNE} we can check that the central charge takes the correct value
\begin{equation}
c=c_{\eta}+c_{\delta}+c_{\varphi}=-2+(1+6k)+1=-9~.
\end{equation}
Indeed, the fact that this free field construction realizes the simple VOA follows from the fact that $T=T_{\text{Sugawara}}$ identically in this construction.

A remark is in order. This free field realization is almost identical to one that was presented in \cite{Adamovic:2014lra} and recently generalized in \cite{Bonetti:2018fqz}.\footnote{
The explicit identification proceeds as follows. The starting point is given in \cite{Adamovic:2014lra} where the $(\beta,\gamma,b,c)$ system is bosonized in terms of three chiral bosons that we denote by $\alpha_{\!\text{ \cite{Adamovic:2014lra}}}$, $\beta_{\!\text{ \cite{Adamovic:2014lra}}}$, $\delta_{\!\text{ \cite{Adamovic:2014lra}}}$. The next step is to defined new chiral bosons as linear combinations as these three as in \cite{Adamovic:2017} with $k=-\frac{3}{2}$. The latter are denoted as $\mu_{\!\text{ \cite{Adamovic:2017}}}$, $\nu_{\!\text{ \cite{Adamovic:2017}}}$, 
$\gamma_{\!\text{ \cite{Adamovic:2017}}}$. With this done, the free field realization above can be identified by setting
\begin{equation}
\delta=-\tfrac{2}{k}\nu_{\!\text{ \cite{Adamovic:2017}}}~,\qquad
\varphi=+\tfrac{2}{k}\mu_{\!\text{ \cite{Adamovic:2017}}}~,\qquad
\eta_1=e^{-\frac{1}{2}\gamma_{\!\text{ \cite{Adamovic:2017}}}}~,\qquad
\eta_2=\partial\gamma_{\!\text{ \cite{Adamovic:2017}}}e^{+\frac{1}{2}\gamma_{\!\text{ \cite{Adamovic:2017}}}}~.
\end{equation}
} The version given here has several advantages. The first is that it is more flexible than the one of \cite{Adamovic:2014lra}. As in the example of Section \ref{subsec:A1_Deligne}, where the existence of a free field realization involving only two chiral bosons singles out three possible values of $k$ (see \eqref{valuesofkA1}) also in this case a free field construction exists for multiple (in fact two) values of $k$. In this case the second value is $k=-\tfrac{1}{2}$ and the free field realization is given by
\begin{equation}
f(z)\,=\,-1\otimes \left(\tfrac{1}{16} (\partial \delta)^2+\tfrac{1}{8}\partial^2\delta \right)e^{-(\delta+\varphi)}\qquad
\begin{pmatrix}G^-(z)\\\widetilde{G}^-(z)\end{pmatrix}\,=\,
\frac{1}{4}\,
\begin{pmatrix}\eta_1\\\eta_2\end{pmatrix}\otimes \partial \delta\,e^{-\frac{\delta+\varphi}{2}}~.
\end{equation}
while the generators \eqref{positiveGeneratorsN4} remains unchanged. The physical interpretation of this second possibility is transparent: this is the VOA associated to the $\mathbb{Z}_2$ quotient of $\mathcal{N}=4$ SYM with gauge group $U(1)$. Another important and useful feature of this realization is that the $SL(2)$ outer  automorphism of the small $\mathcal{N}=4$ super-Virasoro algebra is completely manifest and consists of rotations of $\eta_1$ and $\eta_2$. Both features appear to be very useful in generalization of this construction to higher-rank $\mathcal{N}=4$ theories and higher-genus class $\mathcal{S}$ theories.

\section{\texorpdfstring{$(A_1,D_{\rm odd})$}{(A1,D-odd)} Argyres-Douglas theories}
\label{sec_AD}

Our final family of examples are such that relevant VOA $\mathcal{V}$ is realized in terms of free fields together with an ``irreducible'' building block consisting of a $C_2$-cofinite VOA $\mathcal{V}_{\text{IR}}$, where $\mathcal{V}_{\text{IR}}=\chi[\mathcal{T}_{\text{IR}}]$ is associated to the decoupled interacting theory $\mathcal{T}_{IR}$ that survives at generic points on the Higgs branch. This surviving theory necessarily has trivial Higgs branch, so by \eqref{eqn:M=X} the VOA $\mathcal{V}_{\text{IR}}$ is $C_2$-cofinite. Interestingly, the relevant VOA construction has appeared recently in the mathematical literature \cite{Adamovic:2017}.

The theories in question are the $(A_1,D_{2n+1})$ Argyres-Douglas SCFTs, all of which share the simple Higgs branch
\begin{equation}
\mathcal{M}_{\text{Higgs}}[\text{AD}_{(A_1,D_{2n+1})}]\,=\,\mathbb{C}^2/\mathbb{Z}_2~.
\end{equation}
This is the same Higgs branch we have met several times before, but now at a generic point on the Higgs branch the effective theory, aside from a free hypermultiplet describing the flat directions, includes a decoupled copy of the $(A_1,A_{2n-2})$ Argyres-Douglas SCFT, which has trivial Higgs branch itself.

The VOAs associated to the these Argyres-Douglas theories have been understood to be given by \cite{rastelli_harvard,Cordova:2015nma}
\begin{equation}\label{VandVIRArgyres}
\mathcal{V}= \chi[\text{AD}_{(A_1,D_{2n+1})}]\,=\,V_{-2+\frac{2}{2n+1}}(\mathfrak{sl}(2))~,\qquad
\mathcal{V}_{\text{IR}}=\chi[\text{AD}_{(A_1,A_{2n-2})}]\,=\,\text{Vir}_{2,2n+1}~,
\end{equation}
where $\text{Vir}_{2,2n+1}$ denotes the Virasoro VOA underlying the non-unitary $(2,2n+1)$ minimal model, and the as in the DC series the UV and IR VOAs are related by quantum Drinfel'd-Sokolov reduction.\footnote{In general, DS reduction of $V_{k}(\mathfrak{sl}(2))$ at admissible levels $k+2=\tfrac{p}{q}$ gives the $(p,q)$ minimal model Virasoro algebra. A simple consistency check of this statement is that
\begin{equation}
c_{\text{DS}}\,=\,1-6\frac{\,(k+1)^2}{k+2}\,=\,1-6 \frac{(p-q)^2}{p q}~,\qquad k+2=\frac{p}{q}~.
\end{equation}
} 
According to our general philosophy, the VOA $\mathcal{V}$ should realized as a subVOA of a tensor product of chiral bosons with $\mathcal{V}_{\text{IR}}$,
\begin{equation}
\mathcal{V}\,\subset\,\mathcal{V}_{\text{IR}}\,\otimes \Pi~,
\end{equation}
where $\Pi$ is the lattice subVOA introduced in \eqref{PilattexVOA}. The generators of $\mathcal{V}$ can be found with the same form as in \eqref{eSL2},\eqref{htheta}, and \eqref{fthetaDELIGNE}, but with an important difference. They are given by
\begin{subequations}\label{ADgen}
\begin{align}
e(z)\,&=\,1\otimes\,e^{\delta+\varphi}~,\\
h(z)\,&=\,1\otimes\,k\,\partial\varphi~,\\
\label{f_AD}
f(z)\,&=\,\Big((k+2)T_{\text{IR}}\otimes 1-1\otimes\left((\tfrac{k}{2}\,\partial\delta )^2-\tfrac{k(k+1)}{2}\partial^2 \delta\right)\Big)\Big(1\otimes e^{-(\delta+\varphi)}\Big)~,
\end{align}
\end{subequations}
where $\langle\varphi,\varphi\rangle=-\langle \delta,\delta\rangle=\tfrac{2}{k}$ and $k=-4n/(2n+1)$. Here $T_{\text{IR}}$ is the generator of the (simple) Virasoro VOA at central charge $c(2,2n+1)$. Importantly, this realization has the property that the null states of the VOA $\mathcal{V}$ are proportional to the null operators of the VOA $\mathcal{V}_{\text{IR}}$. In both cases there is a unique singular vector that generate the maximal ideal of the corresponding universal vertex algebra so that to check this statement it is sufficient to verify that it holds for these unique singular vectors. Let us illustrate this point in more detail for the first couple of examples.

First, let us look at the Sugawara stress tensor $T_{\text{Sug}}$ in the free field realization. Direct computation shows that
\begin{equation}
T_{\text{Sug}}:=\tfrac{1}{2(k+2)}\left(ef+fe+\tfrac{1}{2}hh\right)\,=\,T_{\delta}+ T_{\varphi}+ T_{\text{IR}}\,,
\end{equation}
where
\begin{equation}
T_{\delta}+ T_{\varphi}=-\tfrac{k}{4}\left(\partial \delta\, \partial \delta - \partial \varphi \,\partial\varphi\right)+\tfrac{k}{2}\partial^2 \delta~.
\end{equation}
Now we consider the composite operator
\begin{equation}\label{O1def}
\mathcal{O}_1\,\colonequals e\,T_{\text{Sug}}-\tfrac{1}{2}\left(e^\prime h-eh^\prime-\tfrac{1}{3}e^{\prime\prime}\right)~.
\end{equation}
This operator is quasiprimary and highest-weight with respect to $\mathfrak{sl}(2)$, where the latter is the global part of the affine current algebra. In the free field realization \eqref{ADgen}, this composite takes the form
\begin{equation}\label{O1freefields}
\mathcal{O}_1\,=\, \Big(T_{\text{IR}}\otimes 1-1\otimes\left(\tfrac{1}{3}(1+\tfrac{3\,k}{4})\,(\upsilon_+^2-2\,\partial \upsilon_+^{})\right)\Big)\Big(1\otimes e^{\delta+\varphi}\Big)~,
\end{equation}
where $\upsilon_+=\partial(\delta+\varphi)$. The composite operator $\mathcal{O}_1$ defined in \eqref{O1def} is null precisely when the level is given by $k=-\tfrac{4}{3}$. This is easily seen in the free field realization: for $k=-\tfrac{4}{3}$, the expression \eqref{O1freefields} reduces to $\mathcal{O}_1=T_{\text{IR}}\otimes e^{\delta+\varphi}$, moreover for this level, corresponding to $n=1$ in the notation introduced in \eqref{VandVIRArgyres}, the infrared VOA $\mathcal{V}_{\text{IR}}$ reduces to the trivial VOA whose only generator is the identity operator, in particular  $T_{\text{IR}}=0$.  The operator $\mathcal{O}_1$ is thus manifestly zero for this level.

The previous example is clearly exceptional since the IR VOA in this case is trivial. The next case, with $n=2$, presents more structure. To study this case we introduce the operator $\mathcal{O}_2$ in $\mathcal{V}$ which, in terms of free fields, is given by
\begin{equation}\label{O2freefields}
\mathcal{O}_2=\Big(\Lambda_{\text{IR}}\otimes 1+(8+5k)\left(T_{IR}^{}\otimes F_1+T_{IR}^{\prime}\otimes(-\tfrac{5}{32}\upsilon_+)+T_{IR}^{\prime\prime}\otimes\tfrac{1}{32}+1\otimes (4+3k)F_2\right)\Big)\Big(1\otimes e^{\delta+\varphi}\Big)~,
\end{equation}
where
\begin{equation}
\Lambda_{\text{IR}}=(T_{IR}T_{IR})-\tfrac{3}{10}\,T_{IR}^{\prime\prime}~,
\end{equation}
and $F_{1,2}=F_{1,2}[\upsilon_+]$ are polynomials in $\upsilon_+$ and its derivative. Though it will not be important to have the explicit form of this operator in terms of $e(z)$, $h(z)$, and $f(z)$, it is essential (and true) that it is indeed possible to write $\mathcal{O}_2$ entirely in terms of the affine currents. When $k=-\tfrac{8}{5}$ (corresponding to $n=2$) the expression \eqref{O2freefields} reduces to 
\begin{equation}
\mathcal{O}_2=\Lambda_{\text{IR}}\otimes e^{\delta+\varphi}=\Lambda_{\rm IR}\otimes e~.
\end{equation}
In this case, $T_{IR}$ generates the simple Virasoro VOA with central charge $c=c(2,5)=-\frac{22}{5}$, and for this value of the central charge the composite operator $\Lambda_{\text{IR}}$ is null (and thus zero in the simple VOA). Consequently the operator $\mathcal{O}_2$ is identically zero in this free field realization. The cases of higher $n$ work similarly, with an operator $\mathcal{O}_n$ in the affine current VOA becoming proportional to $(T_{IR}^n+\dots)\otimes e^{\delta+\varphi}$ at the appropriate level, with $(T_{IR}^n+\dots)$ the unique singular vector of the relevant Virasoro VOA.

\section{The \texorpdfstring{$R$}{R}-filtration}
\label{sec_filtration}

We have seen that the free field realizations arising from the Higgs branch EFT of a four-dimensional SCFT come equipped with natural $R$-filtrations that we propose (as in \cite{Bonetti:2018fqz} for a different class of theories) to identify with the physical $R$-filtration inherited from four dimensions. Having access to this extra structure is of vital importance in promoting VOA spectral data to four dimensional SCFT data. 

This proposal is motivated by the effective field theory interpretation of the free fields: the chiral bosons that encode the geometry of the Higgs branch inherit a filtration that that arises manifestly from the scaling symmetry of the Higgs branch. Any extra degrees of freedom associated to ``irreducible'' (or $C_2$-cofinite) building blocks must be endowed with their own $R$-filtration.\footnote{This point shows that the characterization of the $R$-filtration for $C_2$-cofinite VOAs related to four-dimensional SCFTs is of particular interest, and this is a subject that we intend to pursue in future work.} Then since the relevant VOA $\mathcal{V}$ is realized as a subVOA of the ``effective field theory'' VOA $\mathcal{V}_{\text{EFT}}$,
\begin{equation}
\mathcal{V}\,\subset\,\mathcal{V}_{\text{EFT}}~,
\end{equation}
having the filtration at the level of the EFT building blocks implies an $R$-filtration on the UV VOA. 

\vspace{1mm}
\begingroup
\begin{table}[t]
\centering
\renewcommand{\arraystretch}{1.4}
\begin{tabular}{| c||c | c |c | c|c |c |c|c  | }
\hline
& $\xi$ & $\partial(\delta-\varphi)$ & $e^{\frac{n}{2}(\delta+\varphi)}$ & $\partial(\delta+\varphi)$ & $\eta_1$ & $\eta_2$ & $T_{IR}$  & $\partial$ \\
\hline \hline
$R$ & $\frac 12$ &$1$ & $\frac n2$ & $0$& $\frac 12$ & $\frac 12$ & $1$ & $0$ \\
\hline
$h-R$ & $0$ &  $0$ &  $0$& $1$ & $\frac 12$ & $\frac 12$ &$ 1$ & $1$ \\
\hline
$r$ & $0$ &  $0$ & $0$ & $0$ &  $+\frac 12$ & $-\frac 12$ & $0$& $0$ \\
\hline
\hline
$h-R-r$ & $0$ &  $0$ &  $0$ &$1$ & $0$ & $1$ & $1$ & $1$ \\
\hline
$h-R+r$ &  $0$ &  $0$& $0$& $1$ & $1$ & $0$ &  $1$ & $1$ \\
\hline
\end{tabular}
\label{tableRweightsLetters}
\end{table}
\endgroup

The $R$-weights assigned to the free fields and $C_2$-cofinite piece appearing in the examples considered in this work are summarized in Table~\ref{tableRweightsLetters}, where $(\delta,\varphi)$ are chiral bosons, $\xi$ are symplectic bosons, and $\eta$ are symplectic fermions. When $\mathcal{V}_{\text{EFT}}$ includes $C_2$-cofinite building blocks, like the one generated by the stress tensor $T_{IR}$ in the previous section, the $R$-filtration of this part should be given independently or determined by other methods.

There are two fundamentally attractive aspects of this proposal. The first is the manifest property that the subspace (of the associated graded ${\rm Gr}_F\VV$) with $h=R$ coincides with the Higgs branch chiral ring. As we emphasized in Section \ref{sec:affinization_and_free}, this fact also provides an efficient way to determine the explicit form of the free field realization by computing ``quantum corrections'' to the expression for the Higgs branch generators by subleading terms in the filtration. More generally, the Hall-Littlewood chiral ring is easily extracted by restricting to operators with $h=R+r$.

The second important non-trivial consistency check of our proposal is that the $R$-filtration arising from free fields assigns weight $R=1$ to the stress tensor.\footnote{Here by assigning weight, we mean that this is the smallest $R$ for which the operator belongs to $F_R\VV$. A more elaborate procedure, which we have not addressed in the present work, would involve using the non-degeneracy of the inner product defined by VOA two-point functions in order to upgrade the filtration to an actual grading on $\VV$. The present notion of $R$-charge assignment for the stress tensor will then match that true $R$-grading assignment.} Recall that in the examples we have analyzed, the stress tensor has taken the form\footnote{Additionally, in the examples we have considered, $T$ coincides with the Sugawara stress tensor so that it is not a strong generator of the VOA.}
\begin{equation}
T=T_{\delta}+T_{\varphi}+T_{\text{rest}}~,\qquad
T_{\text{rest}}\,\in\,\{0,T_{\xi},T_{\eta},T_{\text{IR}}\}=\{0,\,\xi \partial\xi,\,\eta \eta,\,T_{\text{IR}}\}~.
\end{equation} 
On the right hand side of the second equation we have given a schematic form of the corresponding operator that correctly reflects the components of the $R$-filtration to which it belongs. The precise expressions for $T_{\xi}$ and $T_{\eta}$ can be found in and below \eqref{StresstensorTDC} and in \eqref{TetaN=4}, respectively, while $T_{\text{IR}}$ corresponds to the $C_2$-cofinite piece appearing in \eqref{f_AD}. The $R$-weight assigned to $T_{\text{rest}}$ by the rules of Table~\ref{tableRweightsLetters} is manifestly $1$ as it should be. Concerning the remaining piece of the stress tensor, there is a non-trivial cancellation that gives
\begin{equation}
T_{\delta}+T_{\varphi}=-\tfrac{k}{4}\big((\partial \delta)^2-2\partial^2\delta -(\partial \varphi)^2\big)=\tfrac{1}{2}\big(\upsilon_+\upsilon_--\partial \upsilon_-\big)+\tfrac{k}{2}\,\partial \upsilon_+~,
\end{equation}
where 
\begin{equation}\label{upsilondefs}
\upsilon_+\,=\,\partial\left(\delta+\varphi\right)~,\qquad
\upsilon_-\,=\,-\tfrac{k}{2}\,\partial\left(\delta-\varphi\right)~.
\end{equation}
The rules of Table~\ref{tableRweightsLetters} then assign charges $R[\upsilon_+]=0$, $R[\upsilon_-]=1$ so that $R[T_{\delta}+T_{\varphi}]=1$. Notice that this phenomenon follows from the requirement that $e^{\delta+\varphi}$ has regular OPE with itself so that $\delta+\varphi$ is a light-like direction in the lattice of chiral bosons, \ie, $\langle\delta+\varphi,\delta+\varphi \rangle=0$.\footnote{This appear to be the only sensible affine uplift of the $\mathbb{C}^*$ variable $\mathsf{e}$.} Furthermore, the term in $T_{\delta}+T_{\varphi}$ leading in the $R$-filtration, namely $\tfrac{1}{2}\big(\upsilon_+\upsilon_--\partial \upsilon_-\big)$, is entirely determined by the further requirement that $e^{\delta+\varphi}$ has conformal dimension one.

Using the filtration $R$ of the VOA $\mathcal{V}$ one can define the associated graded ${\rm Gr}_R\mathcal{V}$, which is a commutative vertex algebra (in fact a vertex Poisson algebra). This coincides with $\mathcal{V}$ as a vector space, but the multiplication is modified and respects $R$-grading. In general, the associated graded algebra will have more generators than the original VOA. The simplest example of this phenomenon is realized when the stress-tensor of the VOA takes the Sugawara form, which was the case for all examples analyzed in this work. Then the composite operator $\NO{J}{J}|_{\mathfrak{g}\text{-singlet}}\sim T$ has weight $R[T]=1< 2 R[J]=2$. In the associated graded, this produces the (Higgs branch) relation $\mu^2|_{\mathfrak{g}\text{-singlet}}=0$, where $\mu$ is the moment map Higgs branch operator. This leaves the image of the stress tensor in the associated-graded as an independent generator.

This is actually a more general phenomenon: the associated graded has more generators and more relations compared to the original VOA. An interesting example of this phenomenon that does not involve the stress tensor has been found recently in \cite{Buican:2017fiq} and further studied in  \cite{Beem:2019snk}.\footnote{In \cite{Song:2016yfd} a prescription to recover the $R$-filtration has been proposed in some special cases. While extending that prescription to the general situation appears hard and might require intricate knowledge of the underlying four-dimensional theory, the $R$-filtration based on free field realizations is systematically determined after an appropriate (geometric) free field realization is found.} We hope (and expect) our free field realizations to provide a simplified description of the associated graded algebra. The challenge is to find a criterion to determine which elements of ${\rm Gr}_R\mathcal{V}_{\text{EFT}}$ lie in ${\rm Gr}_R\mathcal{V}$. We leave a systematic investigation of this problem for future work.
  
\subsubsection*{Checks against the Macdonald index for \texorpdfstring{$V_{-\frac43}(\slf(2))$}{V(-4/3)sl(2)}}

A rudimentary check for our proposal is that the graded character of the associated-graded of the VOAs in question should match the Macdonald index of the four-dimensional SCFT. Here we will perform some preliminary checks of this claim for the case of the $V_{-\frac43}(\slf(2))$ VOA. The match involves infinite families of operators and is summarized in Table~\ref{table:filtrationExample}.
\begingroup
\renewcommand{\arraystretch}{1.2}
\begin{table}
\begin{center}
\begin{tabular}{|l|c|c|c|c|c|c|}\hline
\diagbox[width=5.5em]{$h-R$}{$h-j$}&
  $0$ &$1$ &$2$ &$3$ &$4$ & $5$    \\ \hline
 \quad\,\,\, \,$0$ & $e^n$ & - & - & - &  -&  -  \\ \hline
 \quad\,\,\, \,$1$ &- & - & $t\,e^n$ & - &- &  - \\ \hline
\quad\,\,\, \,$2$ & - & -  &$e \,\partial^2 e^{n+1}$ &$t\,\partial e^{n+1}$ &$t^2\,e^n$ &  -  \\ \hline
\quad\,\,\, \,$3$ & - & -  &- &\dots &$h\,\partial^3 e^{n+1}$, \, $t\,\partial^2 e^{n+1}$   &  $t^2\,\partial e^{n+1}$ \\ \hline
\quad\,\,\, \,$4$ & - & -  &- &- & \dots &  \dots \\ \hline
\quad\,\,\, \,$5$ & - & -  &- &- & -&  \dots \\ \hline
\end{tabular}
\end{center}
\caption{ 
This table summarizes operators in the VOA $V_{-\frac43}(\slf(2))$, organized according to the two ``twists'' $h-j$ and $h-R$. The label $n$ runs over $0,1,2,\dots,\infty$, and we use the notation $t=T_{\text{Sugawara}}$. Only operators that are both quasi-primaries and $\mathfrak{sl}(2)_F$ highest weight states are displayed. The symbol ``-'' indicates that no operators exist with the given twists and  ``\dots'' indicate additional operators with dimension $h\geqslant6$. The form of the operators is schematic but unambiguous.} 
\label{table:filtrationExample}
\end{table}
\endgroup

The Macdonald index is obtained as a limit of the full superconformal index (see, \eg, \cite{Gadde:2011uv, Rastelli:2014jja}) and depends on two superconformal fugacities, together with generic fugacities for the flavor symmetry which we will leave implicit. It is given by the trace formula
\begin{equation}\label{MacDonaldIndexDEF}
\mathcal{I}_{\rm Macdonald}(q,t): = {\rm STr}_{{\mathcal H}_M  }  (  q^{E - 2 R - r} \, t^{R+r}
   ) \ ,
\end{equation}
where ${\rm STr}$ denotes the supertrace, and ${\mathcal H}_M $ denotes the subspace of the Hilbert space of local operators of the SCFT satisfying $E+2j_1-2R-r=0$. The expression \eqref{MacDonaldIndexDEF} is written in terms of four-dimensional quantum numbers, and the match with the vacuum character of the associated VOA is obtained by recalling that the conformal weight in the chiral algebra is given by $h=E-R$. The explicit expression for the Macdonald index in all the examples studied in this paper is known.\footnote{With the exception of the $\mathfrak{g}_2$ and $\mathfrak{f}_4$ entries of the DC exceptional series, which do not have a known four dimensional origin.} The index for the $\mathfrak{a}_0$ and $\mathfrak{a}_1$ entries of the DC series and the examples of section \ref{sec_AD} can be found in \cite{Song:2016yfd}; the $\mathfrak{a}_2$ entry coincides with $\chi[\text{AD}_{(A_1,D_4)}]$ and its Macdonald index can be found in  \cite{Buican:2015tda}; the $\mathfrak{d}_4$, $\mathfrak{e}_6$, $\mathfrak{e}_7$, and $\mathfrak{e}_8$ entries are class $\mathcal{S}$ theories with regular punctures, so the indices can be obtained using the methods introduced in \cite{Gadde:2011uv}; finally the case of $\mathcal{N}=4$ SYM is reviewed in \cite{Bonetti:2018fqz}. The specific case of interest, $V_{-\frac43}(\slf(2))$, is the associated VOA of the four dimensional Argyres-Douglas theory of type $(A_1,D_3)$, and the Macdonald index can be found in \cite{Song:2016yfd}.
 
Let us proceed by explaining how the Macdonald index can be computed according to our prescription using free fields. The free fields in this example are those of the VOA $\Pi$ defined in \eqref{PilattexVOA}. We recall the expressions for the generators of $\mathcal{V}$ and the Sugawara stress tensor, organized in order of decreasing $R$ assignment.
\begin{align}\label{efhsummary}
e(z)\,&=\,e^{\delta+\varphi}~,\\
\label{efhsummary1}
h(z)\,&=\,\,\upsilon_-\,-\,\tfrac{2}{3}\,\upsilon_+~,\\
\label{efhsummary2}
f(z)\,&=\,\Big(
-\tfrac{1}{4}\,\upsilon_-^2
-\tfrac{1}{3}\,(\upsilon_-\upsilon_++\tfrac{1}{2}\,\partial \upsilon_-)
-\tfrac{1}{9}\,(\upsilon_+^2-\partial \upsilon_+)\Big)e^{-(\delta+\varphi)}~,\\
T_{\text{Sug}}(z)=\,T(z)\,&=\,\tfrac{1}{2}\big(\upsilon_+\upsilon_--\partial \upsilon_-\big)-\tfrac{2}{3}\,\partial \upsilon_+~,
\end{align}
where the $\upsilon_{\pm}$ are defined in \eqref{upsilondefs}. 

To determined the refined vacuum character, it is useful to organize the space of composites of the generators \eqref{efhsummary}, \eqref{efhsummary1}, and \eqref{efhsummary2} in irreducible representations of $\mathfrak{sl}(2)_z\oplus \mathfrak{sl}(2)_F$ where $\mathfrak{sl}(2)_z$ denotes the global part of the Virasoro algebra and $\mathfrak{sl}(2)_F$ the global part of the affine algebra. This is useful because $R$ takes a fixed value within each irreducible representation. We denote by $V^{\text{hw}}_{h,j}$ the vector space of highest weights $(h,j)$ with respect to the Cartan generators of $\mathfrak{sl}(2)_z\oplus \mathfrak{sl}(2)_F$. To compute the refined index we can therefore focus on a fixed value of $(h,j)$ and look for a basis of $V^{\text{hw}}_{h,j}$ that minimizes the $R$ assignments for its elements.

The entries of Table~\ref{table:filtrationExample} are obtained by restricting to operators with fixed $h-j$ and arranging them according to their $h-R$ weight, following the prescription just described. This is a simple exercise when $h-j$ is small, but quickly becomes cumbersome for higher values of this quantum number. The results agree with the Macdonald index for all operators shown in Table~\ref{table:filtrationExample}.

\section{Discussion}
\label{sec_conclusions}

In this work we have introduced free field realizations for VOAs associated to four-dimensional SCFTs whose building blocks are in direct correspondence with the low energy degrees of freedom on the Higgs branch. These realizations are highly economical and have the important property that they give rise to a \emph{simple} VOA, with all singular vectors automatically set to zero. We have focused here on three instructive families of examples. In the first family, all low energy degrees of freedom are associated to the geometry of the Higgs branch. Alternatively, in the second and third families there are additional vector multiplets and interacting degrees of freedom, respectively, present at generic points of the Higgs branch.

The case studies presented in this paper have some additional features that simplify their analysis. Firstly, in all cases there is a flavor symmetry $\mathfrak{g}_F$ such that the infrared $R$-symmetry on the Higgs branch is realized as a subalgebra of $\mathfrak{sl}(2)_R\times \mathfrak{g}_F$. Relatedly, these are examples where Higgsing of the UV theory can be performed at the VOA level by quantum Drinfel'd-Sokolov reduction, and the free field realizations we presented can to a certain extent be thought of as inversions of that DS reduction. A second simplifying factor is that in our examples, the only singularity of the Higgs branch is at the origin, which seems to have important consequences for the complexity (or rather, simplicity) of the free field constructions.

We have found, and will report upon in a future publication, a number of cases where free field realizations similar to the ones presented here can be established, but for which one or both of the aforementioned simplifying features is absent. The emerging picture is that the VOA associated to a given theory $\mathcal{T}$ can be realized in terms of the VOA of the theory obtained from $\mathcal{T}$ by (partial) Higgsing at a point $x$ on the Higgs branch, along with a certain number of free fields:
\begin{equation}
\label{eqdiscuss}
\chi[\mathcal{T}]\,\subset\,  \chi[\mathcal{T}_{\text{IR}}]\otimes\,\mathcal{V}_{\text{free}}~,
\qquad
\mathcal{T}_{\text{IR}}=\mathcal{T}\slice{x}~,
\qquad
x\,\in\,\mathcal{M}_{\text{Higgs}}[\mathcal{T}]~.
\end{equation}
Here $\mathcal{T}\slice{x}$ denotes\footnote{The notation  is borrowed from \cite{Tachikawa:2017byo} to suggest the letter 's' for slicing. The Higgsing appearing in \eqref{eqdiscuss} includes the setup discussed in  \cite{Tachikawa:2017byo} but is more general.} the (irreducible) effective theory supported at the point $x$ on the Higgs branch, while $\mathcal{V}_{\text{free}}$ in general includes free bosons and free fermions, with the number of chiral bosons in $\mathcal{V}_{\text{free}}$ being equal to $n=\text{dim}\mathcal{M}_{\text{Higgs}}[\mathcal{T}]-\text{dim}\mathcal{M}_{\text{Higgs}}[\mathcal{T}_{\text{IR}}]$. However, some of these chiral bosons can be re-written as symplectic bosons, denoted by $\xi$ in this paper, so we write $n=2 (m_{\xi}+m_*)$.

The symplectic variety $\mathcal{M}_{\text{Higgs}}[\mathcal{T}]$ admits is in general a stratified symplectic space with a finite number of symplectic leaves. Based on work to date, it seems that the number $m_*$ is equal to the \emph{depth} of the stratification \cite{SjamaarLerman}. In light of this observation, the simple structure of the Higgs branches considered in this paper allows us to find free field realizations with $m_*=1$. When considering Higgsing such that $\mathcal{T}_{\text{IR}}$ itself possesses a non-trivial Higgs branch (meaning that we have not maximally Higgsed) this procedure can be iterated. In certain cases, like the one analyzed in this work, the construction \eqref{eqdiscuss} can be considered as the inverse operation of Drinfel'd-Sokolov reduction. 

The free field constructions given here have a number of appealing applications. The first of these we have explored in a preliminary way in Section \ref{sec_filtration}: the free field construction gives rise to a natural filtration that we have conjectured to be equivalent to the underlying $R$-filtration arising from the $\suf(2)_R$ charge in four dimensions. This is of particular interest if one aims to get a good understanding of constraints arising from the intersection of four-dimensional unitarity with VOA physics. Examples of interesting unitarity bounds arising from only a rudimentary understanding of the $R$-symmetry structure of the associated VOA have appeared in \cite{Beem:2013sza,Lemos:2015orc, Beem:2018duj}, but a more detailed analysis should be possible with good control over this filtration.

Another key application of free field realizations such as those we have presented is to representation theory of VOAs. On general grounds, as these free field realizations are of the simple quotient of the relevant VOA, the remainder of the free field VOA will furnish a (generally reducible) module for this simple quotient. A concrete case where this approach has been worked in detail is the VOA associated to ${\cal N}=4$ SYM theory with gauge group $SU(2)$, which we have discussed in Section \ref{sec_N4}. Adamovic \cite{Adamovic:2014lra} showed that this VOA admits two irreducible modules in category ${\cal O}$ (the vacuum module and another one), and that the $bc \beta \gamma$ free field VOA decomposes as the direct sum of these two modules.

A related open question is whether the free field realizations that we have discussed admit a further specification in terms of screening charges. In this work, we have described the VOA $\chi[ {\cal T}]$ by giving explicit expressions for its strong generators in terms of free fields. Ideally, $\chi[{\cal T}]$ would be equivalently characterized as the kernel of a set of screening charges inside the free field VOA. A concrete realization of this aspiration can again be found in the example of \cite{Adamovic:2014lra}. Such screening charges might admit a geometric interpretation, as global compatibility conditions for free field realizations ``based'' on overlapping open patches of the Higgs branch. In the examples described in this paper, we were able to find approximations of the Higgs branch in terms of a single open patch, which however failed to cover a measure-zero set of points. One may imagine a more systematic approach, where the full Higgs branch is covered with an atlas of charts, such that the conditions of global compatibility are encoded in a set of screening charges. Such a construction would be somewhat reminiscent of the realization of VOAs in terms of curved $\beta \gamma$ systems, though we anticipate that the details would be quite different.

Finally, the most ambitions application of our findings might be to the classification program of ${\cal N}=2$ four-dimensional SCFTs. We are learning that the VOAs associated to $4d$ SCFTs are very special,
and that they might be fully characterized by the effective field theory on the Higgs branch. Combining these new insights with the constraints of four-dimensional unitarity offers a promising blueprint for 
carving out the space of consistent ${\cal N}=2$ theories.


\section*{Acknowledgments}
The authors are grateful to Philip Argyres, Jacques Distler, Simone Giacomelli, Mario Martone, and Wolfger Peelaers for helpful discussions. The work of CB and CM is supported in part by grant \#494786 from the Simons Foundation. The work of LR is supported in part by the NSF grant PHY1620628.

\bibliographystyle{./aux/ytphys}
\bibliography{./aux/refs}

\providecommand{\href}[2]{#2}\begingroup\raggedright\begin{thebibliography}{10}

\bibitem{Beem:2013sza}
C.~Beem, M.~Lemos, P.~Liendo, W.~Peelaers, L.~Rastelli, {\em et al.},
  ``{Infinite Chiral Symmetry in Four Dimensions},''
  \href{http://dx.doi.org/10.1007/s00220-014-2272-x}{{\em Commun.Math.Phys.}
  {\bfseries 336} no.~3, (2015) 1359--1433},
\href{http://arxiv.org/abs/1312.5344}{{\ttfamily arXiv:1312.5344 [hep-th]}}.

\bibitem{Beem:2014rza}
C.~Beem, W.~Peelaers, L.~Rastelli, and B.~C. van Rees, ``{Chiral algebras of
  class S},'' \href{http://dx.doi.org/10.1007/JHEP05(2015)020}{{\em JHEP}
  {\bfseries 05} (2015) 020},
\href{http://arxiv.org/abs/1408.6522}{{\ttfamily arXiv:1408.6522 [hep-th]}}.

\bibitem{Arakawa:2018egx}
T.~Arakawa, ``{Chiral algebras of class $\mathcal{S}$ and Moore-Tachikawa
  symplectic varieties},''
\href{http://arxiv.org/abs/1811.01577}{{\ttfamily arXiv:1811.01577 [math.RT]}}.

\bibitem{Liendo:2015ofa}
P.~Liendo, I.~Ramirez, and J.~Seo, ``{Stress-tensor OPE in $ \mathcal{N}=2 $
  superconformal theories},''
  \href{http://dx.doi.org/10.1007/JHEP02(2016)019}{{\em JHEP} {\bfseries 02}
  (2016) 019},
\href{http://arxiv.org/abs/1509.00033}{{\ttfamily arXiv:1509.00033 [hep-th]}}.

\bibitem{Lemos:2015orc}
M.~Lemos and P.~Liendo, ``{$\mathcal{N}=2$ central charge bounds from $2d$
  chiral algebras},'' \href{http://dx.doi.org/10.1007/JHEP04(2016)004}{{\em
  JHEP} {\bfseries 04} (2016) 004},
\href{http://arxiv.org/abs/1511.07449}{{\ttfamily arXiv:1511.07449 [hep-th]}}.

\bibitem{Beem:2018duj}
C.~Beem, ``{Flavor symmetries and unitarity bounds in ${\mathcal N}=2$
  SCFTs},''
\href{http://arxiv.org/abs/1812.06099}{{\ttfamily arXiv:1812.06099 [hep-th]}}.

\bibitem{Beem:2017ooy}
C.~Beem and L.~Rastelli, ``{Vertex operator algebras, Higgs branches, and
  modular differential equations},''
\href{http://arxiv.org/abs/1707.07679}{{\ttfamily arXiv:1707.07679 [hep-th]}}.

\bibitem{Arakawa:2010ni}
T.~Arakawa, ``{Associated varieties of modules over Kac-Moody algebras and
  C(2)-cofiniteness of W-algebras},''
\href{http://arxiv.org/abs/1004.1554}{{\ttfamily arXiv:1004.1554 [math.QA]}}.

\bibitem{BMRToAppear}
C.~Beem, C.~Meneghelli, and L.~Rastelli, ``work in progress.''.

\bibitem{Adamovic:2017}
D.~Adamovic, ``{Realizations of simple affine vertex algebras and their
  modules: the cases $\widehat{sl(2)}$ and $\widehat{osp(1, 2)}$},''
  \href{http://arxiv.org/abs/1711.11342}{{\ttfamily arXiv:1711.11342
  [math.QA]}}.

\bibitem{Song:2016yfd}
J.~Song, ``{Macdonald Index and Chiral Algebra},''
\href{http://arxiv.org/abs/1612.08956}{{\ttfamily arXiv:1612.08956 [hep-th]}}.

\bibitem{Bonetti:2018fqz}
F.~Bonetti, C.~Meneghelli, and L.~Rastelli, ``{VOAs labelled by complex
  reflection groups and 4d SCFTs},''
\href{http://arxiv.org/abs/1810.03612}{{\ttfamily arXiv:1810.03612 [hep-th]}}.

\bibitem{Arakawa:2016hkg}
T.~Arakawa and K.~Kawasetsu, ``{Quasi-lisse vertex algebras and modular linear
  differential equations},''
\href{http://arxiv.org/abs/1610.05865}{{\ttfamily arXiv:1610.05865 [math.QA]}}.

\bibitem{Arakawa:2015jya}
T.~Arakawa and A.~Moreau, ``{Joseph ideals and lisse minimal W-algebras},''
\href{http://arxiv.org/abs/1506.00710}{{\ttfamily arXiv:1506.00710 [math.RT]}}.

\bibitem{Dedushenko:2017osi}
M.~Dedushenko and S.~Gukov, ``{A 2d (0,2) appetizer},''
\href{http://arxiv.org/abs/1712.07659}{{\ttfamily arXiv:1712.07659 [hep-th]}}.

\bibitem{Eager:2019zrc}
R.~Eager, G.~Lockhart, and E.~Sharpe, ``{Hidden exceptional symmetry in the
  pure spinor superstring},''
\href{http://arxiv.org/abs/1902.09504}{{\ttfamily arXiv:1902.09504 [hep-th]}}.

\bibitem{Shimizu:2017kzs}
H.~Shimizu, Y.~Tachikawa, and G.~Zafrir, ``{Anomaly matching on the Higgs
  branch},''
\href{http://arxiv.org/abs/1703.01013}{{\ttfamily arXiv:1703.01013 [hep-th]}}.

\bibitem{ArakawaMoreau:Arc}
T.~{Arakawa} and A.~{Moreau}, ``{Arc spaces and chiral symplectic cores},''
  {\em arXiv e-prints} (Feb, 2018) arXiv:1802.06533,
  \href{http://arxiv.org/abs/1802.06533}{{\ttfamily arXiv:1802.06533
  [math.RT]}}.

\bibitem{Adamovic:2004}
D.~Adamovic, ``{A construction of admissible $A_1^{(1)}$-modules of level
  $-4/3$},''
\href{http://arxiv.org/abs/0401023}{{\ttfamily arXiv:0401023 [math]}}.

\bibitem{Semikhatov:1993pr}
A.~M. Semikhatov, ``{The MFF singular vectors in topological conformal
  theories},'' \href{http://dx.doi.org/10.1142/S0217732394001738}{{\em Mod.
  Phys. Lett.} {\bfseries A9} (1994) 1867--1896},
  \href{http://arxiv.org/abs/hep-th/9311180}{{\ttfamily arXiv:hep-th/9311180
  [hep-th]}}.
[JETP Lett.58,860(1993)].

\bibitem{Adamovic:2014lra}
D.~Adamovic, ``{A realization of certain modules for the $N=4$ superconformal
  algebra and the affine Lie algebra $A_2 ^{(1)}$},''
\href{http://arxiv.org/abs/1407.1527}{{\ttfamily arXiv:1407.1527 [math.QA]}}.

\bibitem{Argyres:2007cn}
P.~C. Argyres and N.~Seiberg, ``{S-duality in N=2 supersymmetric gauge
  theories},'' \href{http://dx.doi.org/10.1088/1126-6708/2007/12/088}{{\em
  JHEP} {\bfseries 0712} (2007) 088},
\href{http://arxiv.org/abs/0711.0054}{{\ttfamily arXiv:0711.0054 [hep-th]}}.

\bibitem{ASENS_1976_4_9_1_1_0}
A.~Joseph, ``The minimal orbit in a simple lie algebra and its associated
  maximal ideal,'' \href{http://www.numdam.org/item/ASENS_1976_4_9_1_1_0}{{\em
  Annales scientifiques de l'\'Ecole Normale Sup\'erieure} {\bfseries Ser. 4,
  9} no.~1, (1976) 1--29}.

\bibitem{Kac:1988qc}
V.~G. Kac and M.~Wakimoto, ``{Modular invariant representations of infinite
  dimensional Lie algebras and superalgebras},''
\href{http://dx.doi.org/10.1073/pnas.85.14.4956}{{\em Proc. Nat. Acad. Sci.}
  {\bfseries 85} (1988) 4956--5960}.

\bibitem{arakawa2017sheets}
T.~Arakawa and A.~Moreau, ``Sheets and associated varieties of affine vertex
  algebras,'' {\em Advances in Mathematics} {\bfseries 320} (2017) 157--209.

\bibitem{Gunaydin:2004md}
M.~Gunaydin and O.~Pavlyk, ``{Minimal unitary realizations of exceptional
  U-duality groups and their subgroups as quasiconformal groups},''
  \href{http://dx.doi.org/10.1088/1126-6708/2005/01/019}{{\em JHEP} {\bfseries
  01} (2005) 019},
\href{http://arxiv.org/abs/hep-th/0409272}{{\ttfamily arXiv:hep-th/0409272
  [hep-th]}}.

\bibitem{Joseph:1974hr}
A.~Joseph, ``{Minimal realizations and spectrum generating algebras},''
\href{http://dx.doi.org/10.1007/BF01646204}{{\em Commun. Math. Phys.}
  {\bfseries 36} (1974) 325--338}.

\bibitem{rastelli_harvard}
C.~Beem and L.~Rastelli, ``{Infinite Chiral Symmetry in Four and Six
  Dimensions}.'' Seminar at Harvard University by L. Rastelli, November, 2014.

\bibitem{Cordova:2015nma}
C.~C\'ordova and S.-H. Shao, ``{Schur Indices, BPS Particles, and
  Argyres-Douglas Theories},''
  \href{http://dx.doi.org/10.1007/JHEP01(2016)040}{{\em JHEP} {\bfseries 01}
  (2016) 040},
\href{http://arxiv.org/abs/1506.00265}{{\ttfamily arXiv:1506.00265 [hep-th]}}.

\bibitem{Buican:2017fiq}
M.~Buican, Z.~Laczko, and T.~Nishinaka, ``{N=2 S-duality Revisited},''
\href{http://arxiv.org/abs/1706.03797}{{\ttfamily arXiv:1706.03797 [hep-th]}}.

\bibitem{Beem:2019snk}
C.~Beem, C.~Meneghelli, W.~Peelaers, and L.~Rastelli, ``{VOAs and rank-two
  instanton SCFTs},''
\href{http://arxiv.org/abs/1907.08629}{{\ttfamily arXiv:1907.08629 [hep-th]}}.

\bibitem{Gadde:2011uv}
A.~Gadde, L.~Rastelli, S.~S. Razamat, and W.~Yan, ``{Gauge Theories and
  Macdonald Polynomials},''
  \href{http://dx.doi.org/10.1007/s00220-012-1607-8}{{\em Commun.Math.Phys.}
  {\bfseries 319} (2013) 147--193},
\href{http://arxiv.org/abs/1110.3740}{{\ttfamily arXiv:1110.3740 [hep-th]}}.

\bibitem{Rastelli:2014jja}
L.~Rastelli and S.~S. Razamat,
  \href{http://dx.doi.org/10.1007/978-3-319-18769-3_9}{``{The Superconformal
  Index of Theories of Class $\mathcal {S}$},''} in {\em New Dualities of
  Supersymmetric Gauge Theories}, J.~Teschner, ed., pp.~261--305.
\newblock 2016.
\newblock \href{http://arxiv.org/abs/1412.7131}{{\ttfamily arXiv:1412.7131
  [hep-th]}}.
\newblock
\url{https://inspirehep.net/record/1335343/files/arXiv:1412.7131.pdf}.
\newblock

\bibitem{Buican:2015tda}
M.~Buican and T.~Nishinaka, ``{Argyres-Douglas Theories, the Macdonald Index,
  and an RG Inequality},''
  \href{http://dx.doi.org/10.1007/JHEP02(2016)159}{{\em JHEP} {\bfseries 02}
  (2016) 159},
\href{http://arxiv.org/abs/1509.05402}{{\ttfamily arXiv:1509.05402 [hep-th]}}.

\bibitem{Tachikawa:2017byo}
Y.~Tachikawa, ``{On 'categories' of quantum field theories},'' in {\em
  {Proceedings, International Congress of Mathematicians (ICM 2018): Rio de
  Janeiro, Brazil, August 1-9, 2018}}, pp.~2695--2718.
\newblock 2018.
\newblock
\href{http://arxiv.org/abs/1712.09456}{{\ttfamily arXiv:1712.09456 [math-ph]}}.
\newblock

\bibitem{SjamaarLerman}
R.~Sjamaar and E.~Lerman, ``{Stratified Symplectic Spaces and Reduction},''
  {\em Ann. of Math.} {\bfseries 134} no.~2, (1991) 375--422.

\end{thebibliography}\endgroup

\end{document}